\documentclass[conference]{IEEEtran}
\usepackage[T1]{fontenc}
\usepackage{amssymb}
\usepackage{amscd}
\usepackage{amsmath}
\usepackage{graphicx}
\usepackage{epsfig}
\usepackage{amsthm}
\usepackage{graphics}
\usepackage{psfrag}
\usepackage{rotating}
\usepackage{amsmath} 
\usepackage{amsfonts}
\usepackage{url}
\usepackage{color}
\usepackage{epstopdf}
\usepackage{amsthm}
\usepackage{tikz}
\usetikzlibrary {decorations}
\usepackage{tikz-cd}
\usetikzlibrary {decorations}
\usetikzlibrary{shapes,arrows,intersections, patterns}
\usetikzlibrary{matrix,fit,calc,trees,positioning,arrows,chains,shapes.geometric,shapes,angles,quotes}

\usepackage{algorithm}
\usepackage{algpseudocode}

\usepackage{mathtools}
\usepackage{physics}
\usepackage{multirow}
\usepackage[final]{pdfpages}
\usepackage{enumitem}
\usepackage{multicol}
\usepackage[nolist]{acronym}
\usepackage[font= small]{caption}
\usepackage{subcaption}
\usepackage{ragged2e}
\usepackage{placeins}
\usepackage{array}
\usepackage{multirow}
\usepackage{pgfplots}
\usepackage{pgfplotstable}
\usepackage{calrsfs}
\pgfplotsset{compat=1.12}
\usepgfplotslibrary{polar}
\usepackage{bm}
\usepackage{bbm}
\newcommand{\Nt}{N_{\rm tx}}
\newcommand{\Nr}{N_{\rm rx}}
\newcommand{\Df}{\Delta f}

\newcommand{\Nrfr}{N_{\rm rx}^{\rm rf}}
\newcommand{\Tzero}{T_{\rm 0}}
\newcommand{\rect}{{\sf rect}}
\usepackage[nolist]{acronym}
\usepackage{siunitx}
\DeclareSIUnit{\belmilliwatt}{Bm}
\DeclareSIUnit{\belsquaremeter}{Bsm}
\usepackage{mathtools}

\theoremstyle{remark}
\newtheorem{remark}{Remark}

\DeclareMathAlphabet{\pazocal}{OMS}{zplm}{m}{n}

\newfont{\bbb}{msbm10 scaled 700}

\newfont{\bb}{msbm10 scaled 1100}
\newcommand{\CC}{\mbox{\bb C}}

\newcommand{\RR}{\mbox{\bb R}}

\newcommand{\av}{{\bf a}}
\newcommand{\bv}{{\bf b}}
\newcommand{\cv}{{\bf c}}
\newcommand{\fv}{{\bf f}}

\newcommand{\hv}{{\bf h}}

\newcommand{\sv}{{\bf s}}

\newcommand{\uv}{{\bf u}}
\newcommand{\wv}{{\bf w}}
\newcommand{\vv}{{\bf v}}
\newcommand{\xv}{{\bf x}}
\newcommand{\yv}{{\bf y}}

\newcommand{\zerov}{{\bf 0}}

\newcommand{\Am}{{\bf A}}
\newcommand{\Bm}{{\bf B}}
\newcommand{\Cm}{{\bf C}}
\newcommand{\Dm}{{\bf D}}

\newcommand{\Fm}{{\bf F}}
\newcommand{\Gm}{{\bf G}}

\newcommand{\Id}{{\bf I}}

\newcommand{\Tm}{{\bf T}}
\newcommand{\Um}{{\bf U}}
\newcommand{\Wm}{{\bf W}}
\newcommand{\Vm}{{\bf V}}

\newcommand{\Ym}{{\bf Y}}

\newcommand{\Hc}{{\cal H}}

\newcommand{\Xc}{{\cal X}}

\newcommand{\thetav}{\hbox{\boldmath$\theta$}}

\newcommand{\xiv}{\hbox{\boldmath$\xi$}}

\renewcommand{\det}{{\hbox{det}}}

\renewcommand{\arg}{{\hbox{arg}}}

\renewcommand{\Re}{{\rm Re}}
\renewcommand{\Im}{{\rm Im}}
\newcommand{\eqdef}{\stackrel{\Delta}{=}}

\newcommand{\herm}{{\sf H}}
\newcommand{\trasp}{{\sf T}}
\newcommand{\transp}{{\sf T}}
\renewcommand{\vec}{{\rm vec}}

\newcommand{\Pav}{P_{\rm avg}}
\newcommand{\Na}{N_{\rm a}}
\newcommand{\Nrf}{N_{\rm rf}}
\newcommand{\Ntx}{N_{\rm tx}}

\newcommand{\CBf}{\mathcal{C}_f}

\newcommand{\T}{{\scriptscriptstyle\mathsf{T}}}
\renewcommand{\vec}[1]{{\bm #1}}

\usepackage{hyperref}
\hypersetup{
    colorlinks=true,
    linkcolor=blue,
    filecolor=magenta,      
    urlcolor=blue,
    }
\makeatletter
\def\@tempa#1{\@xp\@tempb\meaning#1\@nil#1}
\def\@tempb#1>#2#3 #4\@nil#5{%
  \@xp\ifx\csname#3\endcsname\mathaccent
    \@tempc#4?"7777\@nil#5%
  \else
    \PackageWarningNoLine{amsmath}{%
      Unable to redefine math accent \string#5}%
  \fi
}
\def\@tempc#1"#2#3#4#5#6\@nil#7{%
  \chardef\@tempd="#3\relax\set@mathaccent\@tempd{#7}{#2}{#4#5}}

\makeatother
\begin{document}

% \title{Super-resolution Parameter Estimation for a beam space OTFS MIMO Radar based on Deep Iterative Unfolding Techniques}
\title{Multi-static Parameter Estimation in the Near/Far Field  Beam Space for Integrated Sensing and Communication Applications}

% %For two column version
% \title{beam space MIMO Radar for Joint  Communication and Sensing with OTFS  Modulation}

% \author{Saeid K. Dehkordi$^{1}$, Jan Hauffen$^{1}$, Peter Jung$^{2}$, Giuseppe Caire$^{1}$
% \thanks{$^1$ Technical University of Berlin, Germany. $^2$ Technical University of Berlin and Fraunhofer HHI , Germany.}
% \thanks{Emails: s.khalilidehkordi@tu-berlin.de}
% }

	\author{\IEEEauthorblockN{
			Saeid K. Dehkordi\IEEEauthorrefmark{1},
		    Lorenzo Pucci\IEEEauthorrefmark{2}, 
			Peter Jung\IEEEauthorrefmark{1}, \IEEEauthorrefmark{3}, 
                Andrea Giorgetti\IEEEauthorrefmark{2},
                Enrico Paolini\IEEEauthorrefmark{2}, 
			Giuseppe Caire\IEEEauthorrefmark{1}}
			
		    \IEEEauthorblockA{\IEEEauthorrefmark{1}Technical University of Berlin, Germany\\
			%\IEEEauthorrefmark{2}Fraunhofer HHI, Berlin, Germany\\
               \IEEEauthorrefmark{2}Wireless Communications Laboratory, CNIT, DEI, University of Bologna, Italy\\
               \IEEEauthorrefmark{3}German Aerospace Center (DLR), Germany\\
			Emails:\{s.khalilidehkordi,peter.jung, caire\}@tu-berlin.de, \{lorenzo.pucci3, andrea.giorgetti, e.paolini\}@unibo.it}}
	
\begin{acronym}
%%% # %%%
\acro{5GB}{$5$th generation and beyond} 
%%% A %%%
\acro{AWGN}{additive white Gaussian noise}
\acro{AoA}{angle of arrival}
\acro{AoD}{angle of departure}
\acro{AMP}{approximate message passing}
%%% B %%%
\acro{BF}{beamforming}
\acro{BS}{Base Station}
%%% C %%%
\acro{CRLB}{Cram\'er-Rao lower bound}
\acro{CFAR}{constant false alarm rate}
%%% D %%%
\acro{DFT}{discrete Fourier transform}
%%% E %%%
\acro{ET}{extended-target}
%%% F %%%
\acro{FMCW}{frequency modulated continuous wave}
\acro{FWHM}{full width at half maximum}
\acro{FoV}{field of view}
\acro{FF}{far-field}
\acro{FISTA}{fast-iterative shrinkage thresholding algorithm}
%%% H %%%
\acro{HDA}{hybrid digital-analog}
\acro{HPBW}{half-power beamwidth}
%%% I %%%
\acro{ISAC}{integrated sensing and communication}
\acro{ISI}{inter-symbol interference}
\acro{ICI}{inter-carrier interference}
\acro{ISTA}{iterative shrinkage thresholding algorithm}
%%% J %%%
\acro{JHi-FISTA}{joint hierarchical - fast iterative soft shrinkage thresholding algorithm}
%%% L %%%
\acro{LoS}{line-of-sight}
\acro{LAMP}{learned approximate message passing}
\acro{LISTA}{learned iterative shrinkage thresholding algorithm}
%%% M %%%
\acro{ML}{maximum likelihood}
\acro{MMV}{multiple measurement vector}
\acro{MIMO}{multiple-input multiple-output}
\acro{mmWave}{millimeter wave}
\acro{MSP}{multi-scatter-point}
%%% N %%%
\acro{NF}{near-field}
%%% O %%%
\acro{OTFS}{orthogonal time frequency space}
\acro{OFDM}{orthogonal frequency division multiplexing}
\acro{OS-CFAR}{ordered statistic constant false alarm rate}
%%% P %%%
\acro{PD}{probability of detection}
\acro{PSD}{power spectral density}
\acro{PSLR}{peak-to-sidelobe ratio}
\acro{PPP}{Poisson point  process}
%%% R %%%
\acro{RF}{radio frequency}
\acro{Rx}{receiver}
\acro{RMSE}{root mean square error}
\acro{RoI}{region of interest}
\acro{RCS}{radar cross section}
%%% S %%%
\acro{sTHZ}{sub-THz}
\acro{SNR}{signal-to-noise ratio}
\acro{SI}{self-interference}
\acro{SLL}{side lobe level}
\acro{SIC}{successive interference cancellation}
\acro{SE}{spectral efficiency}
\acro{SMV}{single measurement vector}
%%% T %%%
\acro{TDD}{time division duplex}
\acro{Tx}{transmitter}
%%% U %%%
\acro{ULA}{uniform linear array}
\acro{UE}{user equipment}
%%% V %%%
\acro{V2V}{vehicle-to-vehicle}
\acro{V2X}{vehicle-to-everything}
\end{acronym}
\maketitle

\vspace{-1.5cm} 

\begin{abstract}
This work proposes a \ac{ML}-based parameter estimation framework for a \ac{mmWave} \ac{ISAC} system in a multi-static configuration using energy-efficient hybrid digital-analog arrays. Due to the typically large arrays deployed in the higher frequency bands to mitigate isotropic path loss, such arrays may operate in the near-field regime. The proposed parameter estimation in this work consists of a two-stage estimation process, where the first stage is based on far-field assumptions, and is used to obtain a first estimate of the target parameters. In cases where the target is determined to be in the near-field of the arrays, a second estimation based on near-field assumptions is carried out to obtain more accurate estimates. In particular, we select \textit{beamfocusing} array weights designed to achieve a constant gain over an extended spatial region and re-estimate the target parameters at the receivers. We evaluate the effectiveness of the proposed framework in numerous scenarios through numerical simulations and demonstrate the impact of the custom-designed flat-gain beamfocusing codewords in increasing the communication performance of the system.      
\end{abstract}

\acresetall
%\vspace{-1cm}


%%%%%%%%%%%%%%%%%%%%%%%%%%%%%%%%%%%%%%%%%%%%%%%%%%%%%%%%%%%%%%%%%%%%%%%%%%%%%%%%%%%%%%%%%%%%%%%%%%%%%%
% The large isotropic path-loss in such frequency bands requires compensation, which is often achieved via highly directional \ac{BF} gain.
\section{Introduction}\label{sec:Introduction}
In the context of 5G and beyond wireless systems, \ac{ISAC} has emerged as one of the key components \cite{ISAC_Survey}. 
Contrary to the active localization already existing in mobile systems, where the \acp{UE} interact with the \acp{BS} for position estimation, the idea behind \ac{ISAC} systems is that of having a wireless communication network composed of one or more \acp{BS}, able to localize passive objects present in the environment only by collecting and analyzing the signal reflected by them, thus using the same hardware resources and the same physical layer used for communication, allowing more efficient use of the spectrum. Incorporating sensing into next-generation mobile systems unlocks vast capabilities for applications like traffic monitoring, pedestrian detection, and urban autonomous driving \cite{liu2022SurveyISAC}.
Recent studies have demonstrated the potential of employing \ac{OFDM}-based waveforms for \ac{ISAC} systems, with a focus in particular on the mono-static configuration, that is, with the \ac{Tx} and \ac{Rx} co-located \cite{braun2010maximum, FullDuplex, Dehk_TWC}. However, this configuration requires extra hardware or digital processing to remove self-interference \cite{FullDuplex}.

A possible solution to avoid the self-interference problem is to resort to a bi-static configuration, where the Tx and Rx are not co-located. Bi-static radar setups are also interesting in that they can extend the sensing area with cost-effective \ac{Rx} units, which may also be mobile \cite{Rappaport_21}. For these reasons, bi-static and multi-static \ac{ISAC} configurations are attracting increasing amounts of interest for future \ac{5GB} networks \cite{SSB_Passive,Bi_static_v2x}.
In order to achieve the required level of delay and angle resolution, radar sensing requires large antenna arrays and wideband signals. Communication waveforms (such as \ac{OFDM}) in \ac{5GB} are expected to use high frequencies (28 — 100 GHz) and large signal bandwidths. Due to the small wavelength, large arrays can be implemented in relatively small form factors. However, the implementation of fully digital architectures becomes extremely challenging due to the enormous data rate of the A/D conversion at each antenna element. In order to alleviate this problem, \ac{HDA} architectures are commonly considered for massive \ac{MIMO} communications. 

Another interesting aspect connected to the use of very large arrays is that, for some users and radar targets, the usual \ac{FF} assumption common to most array processing literature is not satisfied any longer \cite{NF_dardari}. In the case of \ac{NF} propagation, the usual modeling of the received signal as a superposition of planar waves impinging on the array from multiple scattering elements is not valid, and NF-specific algorithm design for communication and sensing is required.
% Wave propagation principles are fundamentally different in \ac{NF} conditions and require specific algorithmic design for communication and sensing.\\ Given the potential scenario that detected objects are \acp{UE}, and considering a practical deployment where the \ac{NF} of the array extends to a distance such that these \acp{UE} could realistically be within that \ac{NF} range, in order to achieve optimal performance of the \ac{ISAC} system, an initial estimate of the operating mode in which the objects (or \acp{UE}) are located is required.

While most of the available literature focuses on the two extreme cases (either the \ac{FF} or \ac{NF} regime \cite{NF_ISAC}) for the algorithmic design of \ac{ISAC} systems, to the best of the authors' knowledge, the state-of-the-art still lacks studies on the sensing and communication performance of a multi-static (or bi-static) \ac{ISAC} system, taking into account both \ac{NF} and \ac{FF} scenarios. In order to employ a beamforming/beamfocusing scheme adapted to the propagation scenario, it is necessary to determine whether the radar target (or the \ac{UE}) is in \ac{FF} or \ac{NF} conditions. This requires some initial estimation, that is agnostic of the propagation conditions. For this reason, this work considers a multi-static \ac{ISAC} configuration composed of two bi-static pairs, capable of detecting and localizing extended sensing objects that are located in the \ac{NF} or \ac{FF} of the \ac{Tx} and/or one of the \ac{Rx} without a priori knowledge.

Motivated by the lower dimensionality of the beamforming codebook and the reduced parameter search space of the \ac{FF} model, the proposed framework \textit{initializes} the sensing operation in the beam-space by using \ac{FF} beamforming vectors and performing \ac{ML} detection and estimation. Based on the target estimation obtained from this initial stage of sensing, the strategy for the second stage is determined. If the presence of targets in the \ac{NF} of the \ac{Tx} is detected, the scheme switches the \ac{Tx} beamforming to a beamfocusing approach, thus increasing the \ac{SNR} at the intended target (UE) location. To obtain the beamfocusing weights, a beam-focusing codebook at \ac{Tx} is utilized where the codewords are designed to maintain an (almost-) constant gain within an extended spatial region. In the communication-only context, requiring some active scheme between the \ac{BS} and \ac{UE}, two-step processes for designing beamformers/beamfocusing have also been advocated in other works (e.g. \cite{Schober_2023}).  Considering the possibility that the detected target may be a communication user, the ability to appropriately illuminate it can lead to a significant increase in the \ac{SNR} and thus an increase in communication performance (i.e. spectral efficiency). In accordance with \cite{Bjorn_Primer, Schober_2023}, our numerical results show that the beamfocusing scheme converges to beamforming within the transitional region of the \ac{NF} and \ac{FF}. 

If the target is detected to be in \ac{FF} of both the \ac{Tx} and \ac{Rx} arrays, a second stage is not required. Nevertheless, as shown in our previous work \cite{Dehk_TWC}, with adequately accurate positional estimates, it is possible to switch to \ac{FF} beamformers with higher directivity to increase the operating \ac{SNR}.
If the target is determined to be in \ac{NF} of an \ac{Rx} array, since the \ac{FF} model used in the first stage is mismatched,  the second stage performs another \ac{ML} estimation of the target parameters with the correct \ac{NF} model, resulting in significantly improved estimation performance. 

%%%%%
The main contributions of this work are highlighted in the following:
\begin{itemize}
\item We consider a multi-static \ac{ISAC} system, composed of one \ac{Tx} and two \acp{Rx}, capable of exploiting the spatial diversity, i.e. the view of the target from different perspectives, to obtain an accurate estimate of extended targets in the system coverage area. Additionally, the setup in this work is based on \ac{HDA} architectures which have been shown to be energy and cost-efficient.
\item We provide a general model, valid for both \ac{NF} and \ac{FF} channel conditions, that accounts for amplitude and phase variations at different antenna elements in the \ac{NF} and converges to the well-known \ac{LoS} propagation model in the \ac{FF}.
\item We propose a two-stage parameter estimation framework that is generalized to the location of the target with regards to being located in the \ac{NF} or \ac{FF} of the multi-static deployment topology.
\item For cases when a target (\ac{UE}) is determined to be in the \ac{NF} of the \ac{Tx}, we introduce a novel algorithm to create a beamfocusing codebook that maintains a constant gain over an extended spatial region. We then demonstrate that it is possible to drastically improve the localization and communication performance by using beam focusing instead of beamforming.
%\item We perform a communication performance analysis to show that it is possible to significantly improve the \ac{SNR}, and thus the channel capacity, by switching from a beamforming approach to a beam focusing approach in the near field.
\item Given that one of the main advantages of multi-static radar setups is to increase system robustness in terms of target detection, we show via numerical examples that by fusing the information from the two \acp{Rx}, it is possible to significantly increase target detection performance.  
\end{itemize}
%%%%
%A well-known advantage of bi-static sensing configuration is to provide an enhanced \ac{RCS} based observation of targets since different viewing perspectives of targets result in different measurements \cite{Griffiths}.  

% The paper is organized as follows. In Section~\ref{sec:}, the considered \ac{ISAC} multi-static system is presented. In particular, the system model, considering both the extended target model and the generalized near/far field channel model, is described. Section~\ref{sec:} presents the \ac{OFDM} input-output relation. Section~\ref{sec:} discusses... In Section~\ref{sec:},..... We then conclude this article with our remarks in Section~\ref{sec:}.
% \subsubsection{Contributions}
% \begin{itemize}
% \item We provide a general model for near and far fielding channel modeling
% \item We provide a two-stage algorithm to estimate Target parameters that \textbf{fuses} the information from the near and far field assumption to improve accuracy
% \item We introduce a novel algorithm to create a beam-focusing codebook for NF communication and sensing.  

% \item mention somewhere in the text that by changing the reduction matrices to  $\Um = \Id_{\Na}$, the derivations apply fro fully digital systems as well.

% \end{itemize}

We adopt the following notations: $(\cdot)^*$ and $(\cdot)^\transp$ denote the complex conjugate and transpose operations, respectively, while $(\cdot)^\herm$ denotes the Hermitian (conjugate and transpose) operation. $\left|x\right|$ denotes the absolute value of $x$ if $x\in\RR$, while $|\Xc|$ denotes the cardinality of a set $\Xc$.  $\|\xv\|_2$ denotes the $\ell_2$-norm of a complex or real vector $\xv$. 
$\Id_m$ denotes the $m \times m$  identity matrix. We let $[n]=\{1, \dots, n\}$ and $[0:n]=\{0, 1, \dots, n\}$ for a positive integer $n$. $\otimes$ denotes Kronecker product.

The rest of the paper is organized as follows. In Section~\ref{sec:phy-model}, the considered extended target model, the near/far field region relationship, and the generalized near/far field channel model are given, while in Section~\ref{IpOpRelation} the \ac{OFDM} input-output relationship for a single bi-static system in the presence of reflections from extended targets is provided. In Section~\ref{MLE_DET} \ac{ML} estimator is derived and our two-stage parameter estimation framework is presented. Section~\ref{BeamFocusing} introduces a novel algorithm that generates a beamfocusing codebook, ensuring a consistent gain across an extended spatial area. Numerical results are presented in Section~\ref{simulation_results}, and Section~\ref{sec:conclusion} concludes the paper with some remarks.

\begin{figure}[h!]
\centering
\begin{subfigure}{0.5\textwidth}
\centering
\scalebox{0.55}{\includegraphics{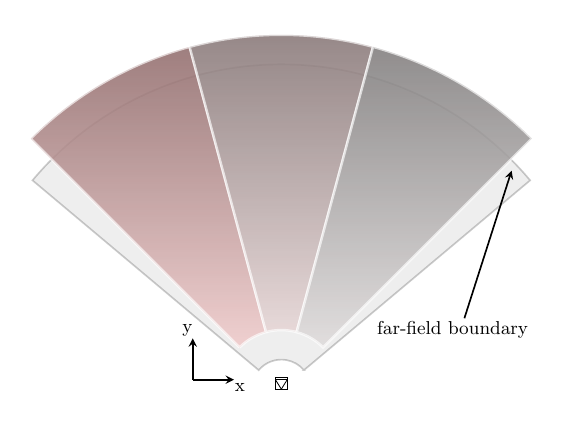}}
\subcaption{}
\label{fig:beamformers_CB} 
\end{subfigure}
%\hfill
%\hspace{0.45cm}
\begin{subfigure}{0.5\textwidth}
\centering
\scalebox{0.55}{\includegraphics{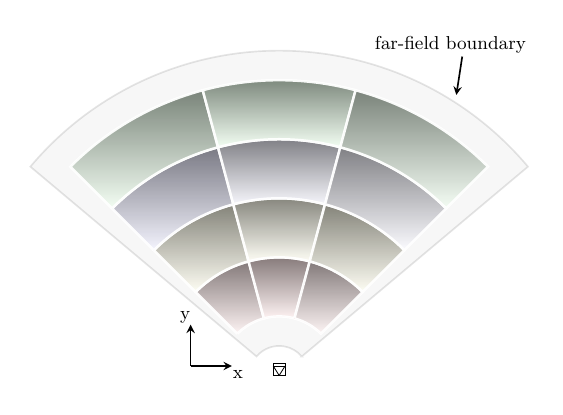}}
\subcaption{}
\label{fig:beam_spots_CB} 
\end{subfigure}
\caption{\small Schematic representation of a codebook of (a) beamforming and (b) beamfocusing codewords in the spatial domain. Note that the \ac{FF} beamformers extend beyond the Fraunhofer distance and beamfocusing codewords are designed to cover only up to the  Fraunhofer distance (i.e. \ac{NF} region). Additionally, the codewords do not need to uniformly divide the space. As an example, in typical urban deployments, areas with more densely located users can be assigned more refined codewords and vice versa.}\label{fig:codebook_scheme}
\end{figure}

%%%%%%%%%%%%%%%%%%%%%%%%%%%%%%%%%%%%%%%%%%%%%%%%%%%%%%%%%%%%%%%%%%%%%%%%%%%%%
\section{System model}\label{sec:phy-model}
In this paper, a multi-static \ac{ISAC} configuration is considered. In particular, as shown in Fig.~\ref{fig:scene_topology}, the system consists of a \ac{Tx} and two \ac{Rx} units to form two bi-static \ac{Tx}-\ac{Rx} pairs.
By using multiple \acp{Rx}, it is possible to see the targets from different perspectives, thus providing a diversity gain, especially in the case of extended targets. In fact, a well-known advantage of bi-static/multi-static sensing configurations is to provide an enhanced \ac{RCS} based observation of targets (i.e., compared to the mono-static configuration), since different observation perspectives of targets result in different measurements \cite{Griffiths}. Another advantage of bi-static setup is that full-duplex processing which would otherwise be required for mono-static sensors is not necessary \footnote{Full-duplex operations can be achieved with sufficient isolation between the transmitter and the radar receiver \cite{sabharwal2014band,duarte2010fullduplexWireless}, nevertheless are still considered a challenge.}. As in most of the related literature, we assume that a connection between the \ac{Tx} and \ac{Rx} units via either a \textit{wired} backhaul or wireless radio link is established.  

In our analysis, we consider \ac{OFDM} as the modulation scheme since it is considered one of the standardized waveforms for 5G-NR \ac{mmWave} systems and \ac{ISAC} applications \footnote{Note that the analysis performed is general and remains valid even if a different multi-carrier modulation scheme is chosen, such as \ac{OTFS}.} \cite{SP-ISAC-Heath-Liu}. In particular, to accomplish the sensing task, we consider that the \ac{Tx} transmits a frame consisting of $N$ \ac{OFDM} symbols, for a duration equal to $N \Tzero$, where $\Tzero \eqdef 1/\Df + T_{\rm cp}$ is the total \ac{OFDM} symbol duration including the cyclic prefix, and with a bandwidth of $W = M \Delta f$, where $M$ is the total number of subcarriers and $\Delta f$ is the subcarrier spacing.
The system operates over a channel with carrier frequency $f_c \gg W$, such that the narrowband array response assumptions hold \cite{VanTrees}.
The generic subcarrier has the frequency $f_m = f_\mathrm{c} + m \Delta f$, where $m = -M/2, \dots, M/2-1$. Aiming at hardware cost and energy efficiency, we consider a \textit{fully-connected} \ac{HDA} array architecture (see e.g.\cite{HDA_Sohrabi}) where the \ac{BS} transmitter is equipped with $\Nrf$ \ac{Tx} \ac{RF} chains driving an antenna array with $\Ntx$ elements. The radar receivers have the same architecture as the \ac{BS}. 

For communication, the \ac{BS} transmits $1 \leq Q \leq \Nrf$ data streams through a beamforming matrix $\Fm = [\fv_1, \dots, \fv_{Q}]$ where $\fv_q$ denotes the $q$-th column of $\Fm$ associated to the $q$-th data stream. The design of \ac{Tx} beamformers $\fv_q$ involves ensuring that each covers a relatively wide section of the beam space with a constant gain, and maintains a very low gain elsewhere (see \cite{Dehk_TWC} for detailed information). In particular, the Tx beamforming vectors are such that $\fv^\herm_{q}\fv_{q'} \approx 0$, for all $q'\neq q$. Subsequently, the backscattered signal originating from the targets (or \acp{UE}) within the beam space segment covered by the respective \ac{Tx} beamforming vectors is utilized for radar processing. It is worth noting that our aim in estimating accurately the \acp{UE} parameters is to eliminate the necessity for active localization feedback in the uplink communication. 

In this work, we assume the number of legitimate targets (i.e. users) is known via communication between the user devices and the \ac{BS}, here serving as the host radar. In this context, the beamforming codewords are selected such that each codeword only covers a single target in the beam space. Therefore, we focus on a scenario where a single data stream (i.e., $\Fm=\fv$) is directed towards the \ac{UE} within a specific sector. 

\begin{remark}
 In this study, in addition to the beamforming codebook used for the \ac{FF} beamforming, the second stage of the scheme comprises \textit{beamfocusing} codewords for transmission in the \ac{NF}. The above assumptions of separated extended targets in the \ac{FF} beam space will be preserved for the \ac{NF} case, where the \ac{Tx} beamfocusing codewords are designed to be non-overlapping and illuminate only a single extended target.   
\end{remark}

The \Ac{mmWave} systems considered for 5G NR and beyond communications are expected to operate under \textit{codebook}-based schemes (e.g. as defined in TS 38.214 \cite{5GNR}). A wide variety of codebook-based schemes have been proposed and investigated in literature (see \cite{Beam_codebook, Hier_CB} and refs therein). This work focuses on the estimation of the spatial parameters of \acp{ET} that are located in a sensing area (Fig.~\ref{fig:scene_topology}) resembling an urban deployment. Given that our approach operates on a codebook basis, we adopt a time division operation mode. In this mode, a \ac{BF} codeword of the \ac{Tx} is selected, and the receivers scan the portion of the beam space illuminated by the respective \ac{Tx} beamformer. Consequently, parameter estimation is carried out specifically within this beam space sector. This process continues sequentially until the entire desired beam space region has been covered. The \ac{HDA} architecture does not allow conventional \ac{MIMO} radar processing, and a vector observation of the beam space (i.e. multiple samples of the beam space) is required for angle estimation. To address this, we define a codebook containing a set of $\Na$ \ac{DFT} orthogonal beams as $\mathcal{U}_{\text{DFT}} \coloneqq (\uv_{1},...,\uv_{\Na}) \in \CC^{\Na\times \Na}$ selected from the Fourier basis ($\in \CC^{\Na\times \Na}$), where $\Na$ is the number of antenna array elements. Subsequently, $\Nrf$ beams out of the $\Na$ are selected at \ac{Rx} units, ensuring coverage of a desired region of interest in the beam space (i.e., covering the illuminated region by the \ac{Tx}) resulting in the formation of the reduction matrix  $\Um \in \CC^{\Na\times \Nrf}$. In scenarios where more than $\Nrf$ beams are required to span the illuminated spatial segment, a multi-block-measurement scheme can be adopted (see \cite{Dehk_TWC,Dehk_WSA} for details). Such a scheme can easily be justified as a result of minimal target movement over the interval  $B$ \ac{OFDM} blocks required for signal acquisition, where $B$ is typically small. 
\begin{remark}
The \ac{HDA} system model in this work can be readily transformed into a fully-digital system. In this case, $\Nrf = \Na$ RF chains are used to demodulate and sample all the antennas of the radar Rx, thus allowing full digital processing.
%by replacing the reduction matrices with  $\Um = \Id_{\Na}$ without loss of generality.
\end{remark}

%%%%%%%%%%%%%%%%%%%%%%%%%%%%%%%%%%%%%%%
%%%%%%%% TARGET MODEL LORENZO %%%%%%%%%
%%%%%%%%%%%%%%%%%%%%%%%%%%%%%%%%%%%%%%%
\subsection{Target model} \label{sec:target_model}
\begin{figure}[t]
\centering
\scalebox{0.5}{\begin{tikzpicture}[auto,rotate=0,transform shape]
\begin{axis}[%
width=2.5in,
height=2.5in,
at={(0in,0in)},
scale only axis,
xmin=-4,
xmax=4,
xlabel style={font=\color{white!10!black},rotate=0},
xlabel={Cross-range},
ymin=0,
ymax=8,
%label style={font=\Large},
%ylabel style={font=\color{white!10!black}},
ylabel={Range},
% axis line style={gray!50!white},
% axis background/.style={fill=white},
%axis x line*=bottom,
%xtick={-5,-4,-2,0,2,4},
xticklabels=\empty,
xticklabel style={rotate=0,xshift=-0.2cm},
% axis y line*=right,
% yticklabel style={rotate=0},
%ytick=\empty,
xmajorgrids,
%xminorgrids,
yticklabels=\empty,
ymajorgrids
]

\begin{scope}[rotate=15]   
    \draw[black, thick, fill=yellow!5, opacity=0.9] (-2,2) rectangle (2,4) node[black!80, right, rotate=0]{};
    \draw[black, thick, step=4mm,opacity=0.8] (-2,2) grid (2,4) node[black!80, right, rotate=0]{};
    
    \draw[black, thick, fill=magenta!95!black, opacity=0.9] (-0.98,3.51) rectangle (-0.5,3.99) node[black!80, right, rotate=0]{};

        \draw[black, thick, fill=magenta!95!black, opacity=0.9] (-1.97,3.51) rectangle (-1.49,3.99) node[black!80, right, rotate=0]{};

        \draw[black, thick, fill=magenta!95!black, opacity=0.9] (-1.97,2.53) rectangle (-1.49,2.99) node[black!80, right, rotate=0]{};

        \draw[black, thick, fill=magenta!95!black, opacity=0.9] (0.04,2.53) rectangle (0.51,2.99) node[black!80, right, rotate=0]{};

        \draw[black, thick, fill=magenta!95!black, opacity=0.9] (1.52,3.51) rectangle (1.995,3.99) node[black!80, right, rotate=0]{};
        \draw[black, thick, fill=magenta!95!black, opacity=0.9] (1.52,2.03) rectangle (2.0,2.51) node[black!80, right, rotate=0]{};
        \draw[black, thick, fill=magenta!95!black, opacity=0.9] (1.02,3.03) rectangle (1.51,3.49) node[black!80, right, rotate=0]{};

        %% Enlarged Grid
        \draw[black, thick, fill=magenta!95!black, opacity=0.9, rotate=-15] (1.5,1) rectangle (2.5,2) node[black!80, right, rotate=0]{};
        \node[rotate=-15] at (axis cs: 1.15,0.2) {T$_g$};
        \node[rotate=-15] at (axis cs: 2.25,1.0) {T$_g$};
        \draw [decorate, rotate=-15, decoration = {brace}] (1.4,1) --  (1.4,2) node[pos=0.5,black]{};
        \draw [decorate, rotate=-15, decoration = {brace}] (1.5,2.1) --  (2.5,2.1) node[pos=0.5,black]{};
        \draw[-latex](2.05,3.75)node[below]{} to[out=20,in=15] (2.8,-0.25);

%%%%%%%%%%%%%%%%%%%%%%%%%%%%%%%%%%%%%%%%%%%%%%%%%%%%%%%

\node[rotate=-15] at (axis cs: 0.1,1.25) {T$_l$};
\node[rotate=-15] at (axis cs: -2.75,2.8) {T$_w$};
\node[] (ant) at (axis cs: -0.5,11) {};
%\node[rotate=60] (txt) at (axis cs: 2.6,10.0) {User antenna};

\node[rotate=-15] at (axis cs: 1.4,6.2) {\small Extended target (observed by Rx$_2$)};
\draw[-latex,thick](-1.0,6)node[below]{} to[out=-140,in=90] (0.15,4.1);

\draw [decorate, decoration = {brace,mirror}] (-2.0,1.8) --  (2,1.8) node[pos=0.5,black]{};
\draw [decorate, decoration = {brace}] (-2.2,2) --  (-2.2,4) node[pos=0.5,black]{};
%\draw [->] (ant) -- (txt);
\end{scope}

\end{axis}

\end{tikzpicture}%
\hspace{1cm}
\hfill

\begin{tikzpicture}[auto,rotate=0,transform shape]
\begin{axis}[%
width=2.5in,
height=2.5in,
at={(0in,0in)},
scale only axis,
xmin=-4,
xmax=4,
xlabel style={font=\color{white!10!black},rotate=0},
xlabel={Cross-range},
ymin=0,
ymax=8,
%label style={font=\Large},
%ylabel style={font=\color{white!10!black}},
ylabel={Range},
% axis line style={gray!50!white},
% axis background/.style={fill=white},
%axis x line*=bottom,
%xtick={-5,-4,-2,0,2,4},
xticklabels=\empty,
xticklabel style={rotate=0,xshift=-0.2cm},
% axis y line*=right,
% yticklabel style={rotate=0},
%ytick=\empty,
xmajorgrids,
%xminorgrids,
yticklabels=\empty,
ymajorgrids
]

   % \draw[black, thick, fill=red!50, step=2mm,opacity=0.3] (-1.8,11.5) grid (1.8,13.2) node[black!80, right, rotate=0]{};

\begin{scope}[rotate=15]   
    \draw[black, thick, fill=yellow!5, opacity=0.9] (-2,2) rectangle (2,4) node[black!80, right, rotate=0]{};
    \draw[black, thick, step=4mm,opacity=0.8] (-2,2) grid (2,4) node[black!80, right, rotate=0]{};
    \draw[black, thick, fill=cyan!95!black, opacity=0.9] (-1.97,2.03) rectangle (-1.49,2.49) node[black!80, right, rotate=0]{};

        \draw[black, thick, fill=cyan!95!black, opacity=0.9] (-1.97,3.03) rectangle (-1.49,3.49) node[black!80, right, rotate=0]{};

        \draw[black, thick, fill=cyan!95!black, opacity=0.9] (-1.48,2.03) rectangle (-1.0,2.49) node[black!80, right, rotate=0]{};
        \draw[black, thick, fill=cyan!95!black, opacity=0.9] (0.06,2.03) rectangle (0.52,2.49) node[black!80, right, rotate=0]{};
        \draw[black, thick, fill=cyan!95!black, opacity=0.9] (0.56,3.03) rectangle (1.02,3.49) node[black!80, right, rotate=0]{};
        \draw[black, thick, fill=cyan!95!black, opacity=0.9] (0.56,3.54) rectangle (1.02,3.97) node[black!80, right, rotate=0]{};
        \draw[black, thick, fill=cyan!95!black, opacity=0.9] (1.05,2.54) rectangle (1.52,3) node[black!80, right, rotate=0]{};

        %% Enlarged Grid
        \draw[black, thick, fill=cyan!95!black, opacity=0.9, rotate=-15] (1.5,1) rectangle (2.5,2) node[black!80, right, rotate=0]{};
        \node[rotate=-15] at (axis cs: 1.15,0.2) {T$_g$};
        \node[rotate=-15] at (axis cs: 2.25,1.0) {T$_g$};
        \draw [decorate, rotate=-15, decoration = {brace}] (1.4,1) --  (1.4,2) node[pos=0.5,black]{};
        \draw [decorate, rotate=-15, decoration = {brace}] (1.5,2.1) --  (2.5,2.1) node[pos=0.5,black]{};
        \draw[-latex](1.55,2.88)node[below]{} to[out=20,in=15] (2.8,-0.25);

    %\draw[step=5mm, black] (2,2) grid (4,5);
   %\node[rectangle,draw, minimum width = 1cm, minimum height = 0.8cm] (r) at (0,11.2) {};
    
  % \begin{scope}[shift={(2.95,8.45)}]    
  %   %\draw(-6,2) node[above]{UE$_2$@ $t_{0}$};
  %   \node[cloud, cloud puffs=5.5, cloud ignores aspect, minimum width=1.5cm, minimum height=1.2cm, align=center, draw] (cloud) at (-1.5cm, 1.8cm) {};
  % \end{scope}  

%%%%%%%%%%%%%%%%%%%%%%%%%%%%%%%%%%%%%%%%%%%%%%%%%%%%%%%

\node[rotate=-15] at (axis cs: 0.1,1.25) {T$_l$};
\node[rotate=-15] at (axis cs: -2.75,2.8) {T$_w$};
\node[] (ant) at (axis cs: -0.5,11) {};
%\node[rotate=60] (txt) at (axis cs: 2.6,10.0) {User antenna};

\node[rotate=-15] at (axis cs: 0.8,6.4) {\small Extended target, area $= |\bf A|$};
\node[rotate=-15] at (axis cs: 0.6,5.8) {\small (observed by Rx$_1$)};
\draw[-latex,thick](-1.0,6)node[below]{} to[out=-140,in=90] (0.15,4.1);

\draw [decorate, decoration = {brace,mirror}] (-2.0,1.8) --  (2,1.8) node[pos=0.5,black]{};
\draw [decorate, decoration = {brace}] (-2.2,2) --  (-2.2,4) node[pos=0.5,black]{};
%\draw [->] (ant) -- (txt);
\end{scope}  

%[decoration={random steps,segment length=2mm}]
%\draw [help lines] grid (3,2);

% \draw [decorate,fill=gray!90!black, opacity=.25]
% (0,0) -- (3,6) arc (0:180:3.2 and 1) -- cycle;

% \draw[opacity = 0.65, pattern=grid, pattern color=gray!90] (2.912,8) -- (5.4596,15) arc [start angle= 70, delta angle= 40, radius=4cm] -- (-2.912,8) arc [start angle= 110, delta angle= -28, radius=2.05cm]-- (2.912,8);
% % r_o = 15.9627
% % r_i = 8.51

% \draw[fill=green!25, opacity = 0.15] (0,0) -- (6.14,16) arc[start angle=69, end angle=110,radius=4.4cm] -- (0,0);

%\draw[-latex,thick](2.0,-2.0)node[below]{Radar Rx$_2$} to[out=90,in=180] (2.9,-2.0);

\end{axis}

% \begin{axis}[%
% width=3.6in,
% height=4.0in,
% at={(0in,0in)},
% scale only axis,
% xmin=0,
% xmax=1,
% ymin=0,
% ymax=1,
% axis line style={draw=none},
% ticks=none,
% axis x line*=bottom,
% axis y line*=left
% ]
% \end{axis}
\end{tikzpicture}%

%https://texample.net/tikz/examples/swan-wave-model/
}
\caption{\small Schematic of the target model, composed of scattering point clusters determined through a BND. Note that each Rx observes a different scattering profile of the extended target at each measurement instance. The parameter values are provided in Tab.~\ref{tab:target-Parameters}. }
\label{fig:PPP_target_model}
\end{figure}
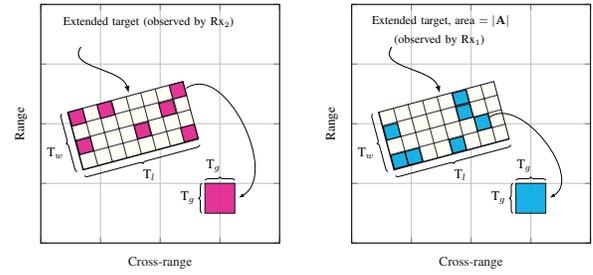

 It is common practice to represent an extended target as a set of fixed point-scatterers. As an alternative approach, it is possible to produce a measurement model (likelihood) in terms of the spatial density of measurements in the intended sensing area. 
Specifically, motivated by finite element discretization techniques commonly used for \ac{RCS} characterization \cite{FEM_1}, in this work, the target is modeled as a set of grid elements $\mathcal{P}$ within a designated rectangular region $\mathbf{A} \subset \RR^2$ with an area of $|\mathbf{A}|$, as shown in Fig.~\ref{fig:PPP_target_model}. At each instant the radar measurement is made, the extended target representing a \ac{UE} (e.g., vehicle, motorcycle, etc.) is composed of a random number $P \leq |\mathcal{P}|$ of scatterers. Given that each grid point inside $\mathbf{A}$ can be active with probability $q$, the number of active points $P$ follows a binomial distribution (BND), with probability $q$ and number of trials $|\mathcal{P}|$, i.e., $P \sim \mathrm{B}(q,|\mathcal{P}|)$. Then, the probability of having $p$ active points at each instant is given by the probability mass function\footnote{Interesting to note that, considering a finite but very large number of elements on the grid, i.e. $|\mathcal{P}| \rightarrow \infty$, each of which is independent active or non-active, the binomial distribution can also be very well approximated by a Poisson distribution, with intensity $\gamma = q |\mathcal{P}|$.}:
\begin{align*}
\mathrm{Pr}(P=p) = \binom{|\mathcal{P}|}{p} q^p (1-q)^{|\mathcal{P}|-p}.
\end{align*}
Then, $P$ points (elements) are drawn i.i.d. from $\mathcal{P}$ such that $P\subseteq \mathcal{P}$.

% Although $P$ is binomial an approximated BND model is considered. 
% Given this, the random variable P can be characterized as a BND with a parameter $\gamma > 0$, and its probability mass function can be defined as:
% \begin{align*}
% \mathrm{Pr}(P=p) = \frac{(\gamma |\bf A|)^{\rm p}}{p!}e^{-\gamma |\bf A|}.
% \end{align*}
% }
In the considered channel model, each individual scattering point (represented by an element) is characterized only by its \ac{LoS} path since \ac{mmWave} channels are characterized by large isotropic attenuation, so multipath components are typically much weaker than the \ac{LoS} and disappear below the noise floor after reflection, especially for the scattering channel seen by the radar receiver (see, e.g., \cite{kumari2018ieee,nguyen2017delay}). With an appropriate choice of grid size, the variable number of scatterers in the target area can be used to model the fluctuations and variance of radar reflectivity of an object caused by target aspect angle, material, etc.\par
Since in this work, we consider a multi-static system, the reflection points observed by each RX unit are generated by a separate BND process at each measurement instance. 

\subsection{Near/far-field region relationship}\label{subsection:NearFar_relation}
\begin{figure}[th!]
\centering
\includegraphics[scale=0.55, angle=0]{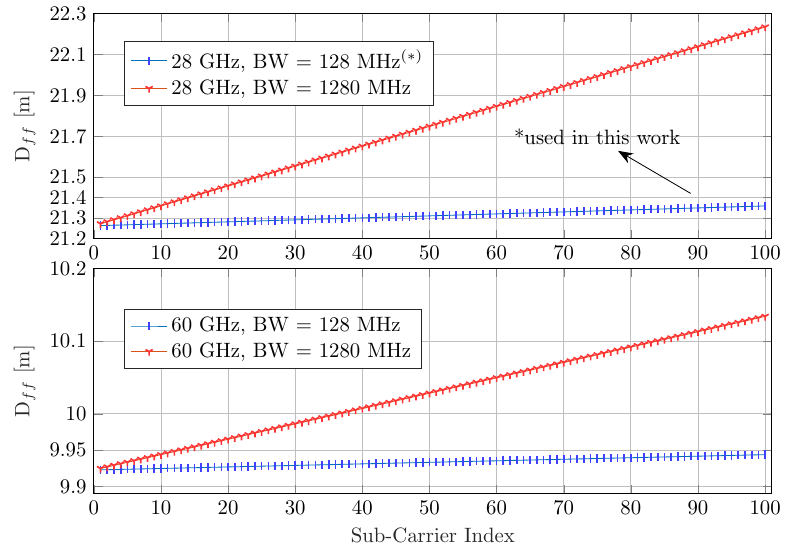}
\caption{FF distance as a function of OFDM sub-carrier index.}
\label{fig:ff_bw}
\end{figure}

\begin{figure*}[ht]
\centering
\includegraphics[scale=0.75, angle=0]{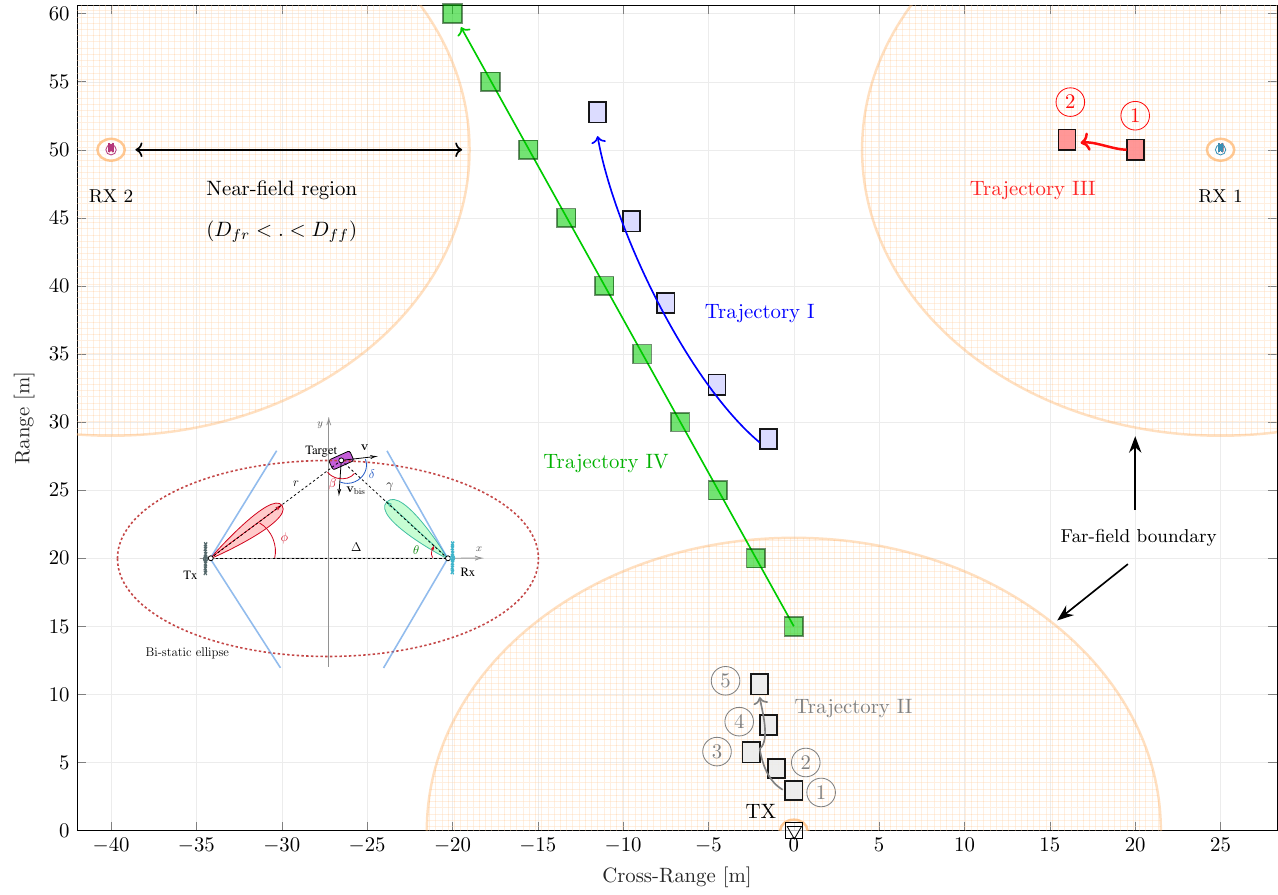}
\caption{System topology resembling an urban deployment scenario. The trajectories indicated in the figure model various locations and movement patterns that may arise with respect to the near and far fields of the antenna arrays. We assume that the deployment has been planned such that the \acp{NF} of the \ac{Tx}/\ac{Rx} pairs do not coincide. The inset depicts the geometric relation between a \ac{Tx}/\ac{Rx} pair and the bi-static ellipse described in Section~\ref{sec:sysmodelRangeDoppler}.}
\label{fig:scene_topology}   
\end{figure*}

The boundary between \ac{NF} and \ac{FF} can
be determined by the \textit{Fraunhofer} distance (also called Rayleigh distance). For an antenna with maximum aperture $D$ at wavelength $\lambda$, the Fraunhofer distance given by $D_{\rm ff}=\frac{2D^2}{\lambda}$ represents the minimum distance for guaranteeing the phase difference of received signals across the array elements of at most $\lambda/4$ \cite{NF_dardari}. For a \ac{ULA} with $\Na$ elements and $\lambda/2$ inter-element spacings , this equates to $\Na^2\lambda/2$. This is widely considered the limit under which wave propagation under the planar assumption holds.
The Fresnel distance $D_{\rm fr}$ given by $\sqrt[3]{\frac{D^4}{8\lambda}}$ is the distance beyond which the reactive field components of the antenna itself become negligible. The distance between $D_{\rm ff}$ and $D_{\rm fr}$ is of interest in this work (see Fig.~\ref{fig:scene_topology}), which is known as the radiative \ac{NF} Fresnel region, or the \ac{NF} region for brevity. 
\begin{remark}
    An important note is in order. When dealing with multi-band systems, the Fraunhofer distance is impacted by the wavelength of each component. More specifically, for the \ac{OFDM} format considered in this work, the Fraunhofer distance varies for each sub-carrier. The overall \ac{FF} regime of the system with a fixed array aperture can then be considered as the Fraunhofer distance of the highest frequency sub-carrier, i.e. $D_\mathrm{ff} = 2D^2/\text{min}(\lambda_m)$. Figure ~\ref{fig:ff_bw} depicts $D_{\rm ff}$ for a few \ac{mmWave} carrier frequencies and bandwidths where the number of \ac{OFDM} subcarriers is $M=100$. As observed, for the system parameters selected in Section~\ref{simulation_results}, the small bandwidth leads to approximately the same $D_{\rm ff}$ across all subcarriers. Nonetheless, for critical cases, this has to be taken into account and the processing needs to be carried out on a per sub-carrier basis separately.    
\end{remark}
%$[2D^2/\text{min}(\lambda_m), \infty)$
\subsection{Channel model}\label{subsection:PhysicalModel}
The considered \ac{ISAC} system can operate in both \ac{NF} and \ac{FF}. In contrast to the \ac{FF} channel model where the signal wavefront is approximated to be a plane, the \ac{NF} channel is modeled to account for a spherical wavefront. This generates a substantial difference between the two regimes. In fact, in \ac{FF} processing, the angle and distance of the target are estimated based on the array response and the time delay of the signal, independently, where the time delay resolution is limited by the system bandwidth. In contrast, in the \ac{NF} regime, it is possible to directly localize the target without estimating time delay, but only by analyzing the phase of the signal scattered by the target and acquired by different antenna array elements, by exploiting the properties of a spherical wave.
However, a general channel model can be found that is valid for both scenarios, i.e. that provides an acceptable approximation in the \ac{NF} and converges as a limit case to the well-known \ac{FF} model as the distance between the target and \ac{Tx}/\ac{Rx} increases.

For simplicity, we consider for a while a single scatterer $p$ (taken from a generic extended target $l$) located at $\mathbf{p}_{p}$ and a transmitting antenna array $k$ whose $N_\mathrm{tx}$ elements are located at $\mathbf{p}_{k_i}$, with index $i=-\frac{N_\mathrm{tx}-1}{2},\dots, \frac{N_\mathrm{tx}-1}{2}$, and an inter-element spacing of $d$. Considering \ac{LoS} propagation conditions, the equivalent low-pass complex channel coefficient for the channel between the single antenna element $k_i$ and the scatterer $p$ at subcarrier $m$ and time $n$ can be written as
\begin{equation}
\label{eq:ch_coeff}
\eta^{(m,n)}_{k_i,p} =  \sqrt{\xi_{k_i,p}}e^{-j \varphi^{(m)}_{k_i,p}} e^{j 2 \pi n T_0 \nu_{k,p}}   
\end{equation}
where $\nu_{k,p}$ is the Doppler shift, $\varphi^{(m)}_{k_i,p}$ is the phase term given by
\begin{equation}
\label{eq:phase}
   \varphi^{(m)}_{k_i,p} = 2 \pi \frac{r_{k_i,p}}{c} f_m + \varphi_0, %-2 \pi n  T_0 f_{\mathrm{D},r} 
%   \nu_{k,l} & = \frac{1}{\lambda_\mathrm{c}}\frac{d}{dt}\left[r_{k,l}(t)\right]
\end{equation}
and $\xi_{k_i,l}$ is the gain factor of the channel between the antenna $k_i$ and the scatterer $p$, which, considering \ac{LoS} propagation conditions, can be written as follows
%%%
\begin{equation}
\xi_{k_i,p} = \frac{\sigma_{k,p}}{4 \pi r_{k_i,p}^2}
\label{eq:tx_gain}
\end{equation}
%%%%
with $r_{k_i,p} = \norm{\mathbf{p}_p - \mathbf{p}_{k_i}}$ the distance between the $k_i$th antenna  of the \ac{Tx} and the scatterer $p$, $c$ is the speed of light,  $\varphi_0 \in \mathcal{U}_{[0, 2\pi)}$ is the phase offset between \ac{Tx} and \ac{Rx}, and $\sigma_{k,p}$ the \ac{RCS} of the scatterer $p$, illuminated by the \ac{Tx} $k$.

As previously stated, the expression in \eqref{eq:ch_coeff} is general and remains valid for both \ac{FF} and \ac{NF} scenarios. Later it will be shown that with some approximations resulting by considering the array size much less than $r_{k,p}$, which is the distance between the center of the transmitting array $k$th (chosen as the reference point) and the scatterer $p$, and the same amplitude at each antenna element of the array, \eqref{eq:ch_coeff} can be simplified in the \ac{FF}. To better highlight the generality of the model, \eqref{eq:phase} can be rewritten as
%%% New channel formalization, more general %%%
%%
\begin{equation}
    \varphi^{(m)}_{k_i,p} = 2\pi \frac{r_{k,p}}{c} f_m + 2\pi \frac{r_{k_i,p} - r_{k,p}}{c} f_m + \varphi_0 
\end{equation}
and as a result, \eqref{eq:ch_coeff} becomes
\begin{align}
\label{eq:ch_coeff_2}
& \eta^{(m,n)}_{k_i,p}  \nonumber\\
& =  \sqrt{\xi_{k_i,p}} \, e^{-j \left (2\pi \frac{r_{k,p}}{c} (f_c + m \Delta f) + 2\pi \frac{r_{k_i,p} - r_{k,p}}{c} f_m + 2 \pi n T_0 \nu_{k,p} + \varphi_0 \right)}.
\end{align}
From \eqref{eq:ch_coeff_2} the channel vector $\boldsymbol{\eta}^{(m,n)}_{k,p} \in \mathbb{C}^{1 \times N_\mathrm{tx}}$  associated with subcarrier $m$ and scatterer $p$ at time $n$, can be obtained as:
\begin{equation}
\label{eq:tx_ch_vector}
\boldsymbol{\eta}^{(m,n)}_{k,p} =  \alpha_{k,p}e^{j 2 \pi (n T_0 \nu_{k,p} - m \Delta f \tau_{k,p})}\mathbf{a}^H(\phi_{k,p}, r_{k,p})
\end{equation}
where $\alpha_{k,p} = \sqrt{\xi_{k,p}}e^{-j (2 \pi f_\mathrm{c}\tau_{k,p} + \varphi_0)}$  is the reference channel coefficient computed with respect to the center of the antenna array, with $\tau_{k,p} = r_{k,p}/c$ the reference propagation delay and $\xi_{k,p}$ the reference channel gain, while $\mathbf{a}(\phi_{k,p}, r_{k,p}) \in \mathbb{C}^{N_\mathrm{tx} \times 1}$, is the array response vector, computed considering $W \ll f_\mathrm{c}$, and defined as:
\begin{align} \label{eq:array_response}
    \av (\phi_{k,p},r_{k,p}) = \left(\begin{array}{c}
    \frac{r_{k,p}}{r_{0,p}} \exp(-j \frac{2 \pi f_c}{c}(r_{0,p} - r_{k,p})) \\
    \frac{r_{k,p}}{r_{1,p}} \exp(-j\frac{2\pi f_c}{c}(r_{1,p}-r_{k,p})) \\ 
         \vdots \\
    \frac{r_{k,p}}{r_{N_\mathrm{tx}-1,l}}\exp(-j\frac{2\pi f_c}{c}(r_{N_\mathrm{tx}-1,l}-r_{k,p})) 
    \end{array}\right).
\end{align}
%
%%%%%%%%%%%%%%%%%%%%%%%%%%%%%%%%%%%%%%%%%%%%%%%%%%%%%%%%%%%%%%%%%%%%%%%%%%%%%%%%
%%%%%%%%%%%%%%%%%%%%%%%%%%%%%%%%%%%%%%%%%%%%%%%%%%%%%%%%%%%%%%%%%%%%%%%%%%%%%%%%
 The relationship between the reference distance $r_{k,p}$ and the distance of the $k_i$th antenna element from the scatterer $p$ can be obtained from a second order Taylor expansion, considering $\mathbf{p}_p= r_{k,p}[\cos{\phi_{k,p}},\, \sin{\phi_{k,p}}]^T$ and $\mathbf{p}_{k_i}=[0,\,id]^T$, as follows
\begin{equation} \label{eq:2ndOrdTaylor}
r_{k_i,p} - r_{k,p} \approx \frac{(i d \cos\phi_{k,p})^2}{2 r_{k,p}} - i d \sin\phi_{k,p}
\end{equation}
%
% %%%%%%%%%%%%%%%%%%%%%%%%%%%%%%%%%%%%%%%%%%%%%%%%%%%%%%%%%%%%%%%%%%%%%%%%%%%%%%%%
% \begin{figure*}[ht]
% \centering
% \includegraphics[scale=0.75, angle=0]{Plots_Figures/Thesis_Scenario.pdf}
% \caption{System topology resembling an urban deployment scenario. The trajectories indicated in the figure model various locations and movement patterns that may arise with respect to the near and far fields of the antenna arrays. We assume that the deployment has been planned such that the \acp{NF} of the \ac{Tx}/\ac{Rx} pairs do not coincide. The inset depicts the geometric relation between a \ac{Tx}/\ac{Rx} pair and the bi-static ellipse described in Section~\ref{sec:sysmodelRangeDoppler} }
% \label{fig:scene_topology}   
% \end{figure*} 
% %%%%%%%%%%%%%%%%%%%%%%%%%%%%%%%%%%%%%%%%%%%%%%%%%%%%%%%%%%%%%%%%%%%%%%%%%%%%%%%%
where $\phi_{k,p}$ is the reference \ac{AoD} related to the scatterer $p$.
By replacing \eqref{eq:2ndOrdTaylor} in \eqref{eq:array_response} and considering $id \ll r_{k,p} \, \forall i$, and $r_{k,p}/r_{k_i,p} \approx 1$ as reasonable assumptions in the \ac{FF}, the array response vector in \eqref{eq:array_response} when $d = c/(2 f_c)$, \eqref{eq:array_response} can be rewritten as
\begin{equation}\label{eq:array_response_FF}
\mathbf{a}(\phi_{k,p}) = [e^{-j \frac{N_\mathrm{tx}-1}{2} \pi \sin \phi_{k,p}}, \dots, e^{j \frac{N_\mathrm{tx}-1}{2} \pi \sin \phi_{k,p}}]^T.
\end{equation}
The channel between a given scatterer $p$ and a \ac{Rx} unit $j$, consisting of $N_\mathrm{rx}$ antenna elements can be modeled in the same way as the one between \ac{Tx} and the scatterer. The only difference is in the channel gain coefficient, as shown below.
In particular, the channel vector $\boldsymbol{\eta}^{(m,n)}_{p,j} \in \mathbb{C}^{ N_\mathrm{rx} \times 1}$ related to the $m$-th subcarrier and $n$-th time instant is given by
\begin{equation}
\label{eq:rx_ch_vector}
\boldsymbol{\eta}^{(m,n)}_{p,j} =  \beta_{p,j}e^{j 2 \pi (n T_0 \nu_{p,j} -m \Delta f \tau_{p,j}})\mathbf{b}(\theta_{p,j}, \gamma_{p,j})
\end{equation}
where $\beta_{p,j} = \sqrt{\zeta_{p,j}}e^{-j 2 \pi f_\mathrm{c}\tau_{p,j}}$ is the reference channel coefficient computed with respect to the center of the $j$-th receiving antenna array, and $\tau_{p,j} = \gamma_{p,j}/c$ the reference propagation delay with $\gamma_{p,j}$ is the distance between scatterer $p$ and the center of antenna array $p$, and $\zeta_{p,j}$ the reference channel gain, while $\mathbf{b}(\theta_{p,j}, \gamma_{p,j}) \in \mathbb{C}^{N_\mathrm{rx} \times 1}$ is the array response vector computed as in \eqref{eq:array_response}, with $\theta_{p,j}$ the reference \ac{AoA}.
The gain factor $\zeta_{p,j}$, considering \ac{LoS} propagation conditions and isotropic antenna elements with effective area $A = c^2/(4 \pi f_\mathrm{c}^2)$, can be written as  
%%%
\begin{equation}
\zeta_{p,j} = \frac{c^2}{(4 \pi \, f_c \, \gamma_{p,j})^2}.
\label{eq:rx_gain}
\end{equation}
%%%

Now, we consider a generic extended target $l$, which is modeled as a group of independent scatterers generated according to a BND, as described in Section~\ref{sec:target_model}. Each of the \ac{Tx}-\ac{Rx} bi-static pairs with index $(k,j)$ can collect echoes from a random set $P_{k,j}^{(l)} \subseteq \mathcal{P}$ of these points. \footnote{In the following, for notation brevity, we indicate with $P = |P_{k,j}^{(l)}|$ the random number of scatterers related to target $l$.} The $1 \times N_\mathrm{tx}$ time-varying channel between a generic \ac{Tx} $k$ and a scattering point $p \in [P_{k,j}]$ is given by \footnote{Note that in the considered time-varying channel models, the Doppler shift term is ignored for ease of notation and will be reintroduced into the model later.}
\begin{equation}
\label{eq:tx_scatt_channel}
\hv_{k,{p}}(t) = \alpha_{k,p} \delta(t-\tau_{k,p}) \mathbf{a}^H(\phi_{k,p}, r_{k,p})
\end{equation}
% where $\gamma_{k,p}$ is the complex attenuation factor, 
where $\mathbf{a}(\phi_{k,p}, r_{k,p}) \in \mathbb{C}^{N_\mathrm{tx} \times 1}$ is the array response vector associated to the scatterer $p$, defined as in \eqref{eq:array_response}, and $\delta(\cdot)$ is the Dirac delta function.

Similarly, the $N_\mathrm{rx} \times 1$ time-varying channel between a generic Rx $j$ and a scattering point $p$ is given by
\begin{equation}
\label{eq:rx_scatt_channel}
\mathbf{h}_{p,j}(t) = \beta_{p,j} \delta(t-\tau_{p,j}) \mathbf{b}(\theta_{p,j}, \gamma_{p,j}).
\end{equation}
Next, considering \eqref{eq:tx_scatt_channel} and \eqref{eq:rx_scatt_channel} and a bi-static pair $(k,j)$, the $P \times N_\mathrm{tx}$ channel between the Tx $k$ and the target $l$, and the $N_\mathrm{rx} \times P$ channel between the Rx $j$ and the target $l$, can be written as 
%%% Channel model with convolution %%%%%
\begin{equation}
\label{eq:1st_channel}
\mathbf{H}_{k,l}(t) = [\mathbf{h}^T_{k,1}(t), \dots, \mathbf{h}^T_{k,P}(t)]^T
\end{equation}
%%%%%
\begin{equation}
\label{eq:2nd_channel}
\mathbf{H}_{l,j}(t) = [\mathbf{h}_{1,j}(t), \dots, \mathbf{h}_{P, j}(t)]
\end{equation}
%%%%%%%%%%%%%
respectively.\\
%in the presence of $L$ extended targets
The $N_\mathrm{rx} \times N_\mathrm{tx}$ time-varying \ac{MIMO} channel can be obtained as the convolution of the two downlink channels in \eqref{eq:1st_channel} and \eqref{eq:2nd_channel}, and is given by
\begin{align}\label{eq:channel}
 & \mathbf{H}_{k,j}(t)  = \mathbf{H}_{l,j}(t) * \mathbf{H}_{k,l}(t) \nonumber\\
  & =  \sum_{p =1}^{P} \varepsilon_{p}  \mathbf{b}(\theta_{p,j},\gamma_{p,j}) \mathbf{a}^H(\phi_{k,p}, r_{k,p}) \delta(t-\tau_{p}) %e^{2j\pi \nu_{p_{i,l}} t} 
\end{align}
%%%
where $\varepsilon_{p} = \alpha_{k,p}\beta_{p,j}$ is the bi-static complex attenuation factor, $\tau_{p} = \tau_{k,p} + \tau_{p,j}$ is the bi-static propagation delay associated with the $p$-th scatterer of the target.
%%%%%%%%%%%%%%%%%%%%%%%%%%%%%%%%%%%%%%%%%%%%%%%%%%%%%%%%%%%%%%%%%%%%%%%%%%%%%%%%
% \begin{figure*}[ht]
% \centering
% \scalebox{1.1}{\input{Plots_Figures/Multi_static_Scenario}}
% \caption{\small Schematic of the target model, composed of scattering point clusters, along with  the \ac{Tx} and \ac{Rx} units arranged in a multi-static setup.}
% \label{fig: multi-static-model}
% \end{figure*}

% \begin{figure*}[h!]
% \centering
% \includegraphics[scale=0.75, angle=0]{Plots_Figures/scene_with_NF_FF.pdf}
% \caption{Scenario and target estimates.}
% \label{fig:box_plot}
% \end{figure*}

% \begin{figure*}
% \centering
%     \resizebox{0.6\textwidth}{!}{
%     \input{Plots_Figures/Multi_static_scenario_v2}
%     }
%     \caption{Schematic of the considered multi-static ISAC system with two Txs and two lRxs. The targets in the monitored area are modeled as clusters of scattering points.}
%    \label{fig:scenario}
% \end{figure*}
%%%%%%%%%%%%%%%%%%%%%%%%%%%%%%%%%%%%%%%%%%%%%%%%%%%%%%%%%%%%%%%%%%%%%%%%%%%%%%%%
% $\hat{\phi}_{k,l}$ is the pointing direction of the \ac{Tx} beamformer to the target $l$,
%%%%%%%%%%%%%%%%%%%%%%%%%%%%%%%%%%%%%%%%%%%%%%%%%%%%%%%%%%%%%%%%%%%%%%%%%%%%%%%%
\section{OFDM Input-Output Relationship}\label{IpOpRelation}
The \ac{OFDM} frame transmitted from \ac{Tx} $k$ to target $l$ is given by
\begin{align}\label{eq:OFDM_tx}
    % \sv_{k,l}(t) = \fv (\hat{\phi}_{k,l}, \hat{r}_{k,l})\sum_{n=0}^{N-1}\sum_{m=0}^{M-1} x_{k,l}[m,n]g_{m, n}(t)
        \sv_{k,l}(t) = \fv_l \sum_{n=0}^{N-1}\sum_{m=0}^{M-1} x_{k,l}[m,n]g_{m, n}(t)
\end{align}
where $\fv_l \in \CC^{\Nt \times 1}$ is a (unit-norm) transmit beamforming (in the \ac{FF}) or beamfocusing (in the \ac{NF}) vector to point toward $l$, and $x_{k,l}[m, n]$ is a generic complex modulation symbol of the $M \times N$ OFDM resource grid used at the \ac{Tx} $k$ to estimate range, angle, and velocity parameters of the target $l$; $ g_{m, n}(t)$ is given by
\begin{align}
\label{eq:pulse_shape}
    g_{m, n}(t) = \rect\left(\frac{t - n\Tzero}{\Tzero}\right)e^{j 2\pi m \Df (t-T_{\rm cp}-n\Tzero)}\,
\end{align}
where $\rect(x)$ is a pulse-shaping function taking value $1$ when $0 \leq x \leq 1$ and $0$ elsewhere.
% \begin{remark}
%    The arguments in the \ac{Tx} beamforming vector $\fv (\hat{\phi}_{k,l}, \hat{r}_{k,l})$ indicate that the beamformer can be steered towards a desired direction (location in the \ac{NF} case). If chosen from a fixed codebook, a different codeword is selected to cover the desired region. In other cases, such as the tracking method in \cite{pedraza2023extended}, a directional beam can be electronically steered based on the tracking prediction from previous measurements.  
% \end{remark}

%%%%%%%%%%%%%%%%%%%%%%%%%%%%%%%%%%%%%%%%%
Next, given the convolution between the time-varying \ac{MIMO} channel in \eqref{eq:channel} and the transmitted signal in \eqref{eq:OFDM_tx}, the noise-free $N_\mathrm{rx}$-dimensional continuous-time signal transmitted by $i$ Tx, scattered by the extended target, and received from the $j$th Rx is given by
\begin{align}
\label{eq:time_rx_signal}
\mathbf{r}_{k,j}(t) = & \sum_{p =1}^{P} \varepsilon_{p}  \mathbf{b}(\theta_{p,j},\gamma_{p,j}) \mathbf{a}^H(\phi_{k,p}, r_{k,p}) \mathbf{s}_{k,l}(t-\tau_{p}) \nonumber \\
= & 
  \sum_{p =1}^{P} \varepsilon_{p}  \mathbf{b}(\theta_{p,j},\gamma_{p,j}) \mathbf{a}^H(\phi_{k,p}, r_{k,p}) \cdot \nonumber \\
   & \fv_l \sum_{n=0}^{N-1}\sum_{m=0}^{M-1} x_{k,l}[m, n]g_{m, n}(t - \tau_{p}).
\end{align}
%%%
As previously mentioned, the \ac{Tx} beamformers/beamfocusers are designed to cover non-overlapping sections in the beam space. In the simultaneous presence of multiple (extended) targets we place an assumption that the $L$ (extended) targets are sufficiently separated in space, i.e., $|\av^H(\phi_{k,l}, r_{k,l})\fv_{l'}(\hat{\phi}_{k,l'},\hat{r}_{k,l'})| \approx 0$ for $l' \neq l$, where $\hat{\phi}_{k,l'}$ is the pointing direction (\ac{AoD}) for a \ac{Tx} beamformer $\fv_{l'}$  associated with the $l'$th (extended) target. 
% Here, in the simultaneous presence of multiple (extended) targets two extreme possibilities arise. 1) Assuming that the $L$ (extended) targets are sufficiently separated in space, i.e., $|\av^H(\phi_{k,l}, r_{k,l})\fv_{l'}(\hat{\phi}_{k,l'},\hat{r}_{k,l'})| \approx 0$ for $l' \neq l$, where $\hat{\phi}_{k,l'}$ is the pointing direction (\ac{AoD}) for a \ac{Tx} beamformer $\fv_{l'}$  associated with the $l'$th (extended) target, then  \eqref{eq:time_rx_signal} can be rewritten as follows
% %%%%%
% \begin{align}
% \mathbf{r}_{k,j}(t) = &\frac{1}{\sqrt{L}} \sum_{l=1}^{L}
%   \sum_{p=1}^{P^{(l)}} \varepsilon_{p} \mathbf{b}(\theta_{p,j},\gamma_{p,j}) \mathbf{a}^H(\phi_{k,p},r_{k,p}) \cdot \nonumber \\ & \cdot \fv_l \sum_{n=0}^{N-1}\sum_{m=0}^{M-1} x_{k,l}[m, n]g_{m, n}(t - \tau_{p}).
% \end{align}
% %%%%%%%%
% 2) Multiple (extended) targets fall in the \ac{FoV} of a given \ac{Tx} beamformer. It is then easy to see that \eqref{eq:time_rx_signal} is the superposition of the scattered signal from all targets within the same beamformer. This model can be extended to a mixture of the two scenarios mentioned above in a straightforward manner.

In the following, we focus on using a single \ac{Tx} beamformer such that multiple segments of the beam space are served in TDM fashion. Aiming at reducing hardware complexity and energy consumption at the radar receiver, we process the received signal $\mathbf{r}_{k,j}(t)$ by a reduction matrix before sampling. In particular, we focus the radar receiver on a single target $l$ for $B$ \ac{OFDM} frames and estimate different targets sequentially in time. To achieve this, a reduction matrix $\Um_{b} \in \CC^{\Nr\times\Nrfr}$ is chosen to cover a particular area in the beam space (determined by the current \ac{Tx} beamformer). 
Then, after the \ac{OFDM} demodulator, considering negligible \ac{ICI} and \ac{ISI} and including noise, a received time-frequency grid of complex elements $y[m, n]$ is obtained at each \ac{RF} chain. The expression of the bi-static wireless channel between \ac{Tx} $k$ and \ac{Rx} $j$, which is the \ac{DFT} sample of \eqref{eq:channel} at $m$-th subcarrier and $n$-th time instant, can be easily obtained as the outer product between the $N_\mathrm{rx} \times 1$ channel vector in \eqref{eq:rx_ch_vector} and the $1 \times N_\mathrm{tx}$ channel vector in \eqref{eq:tx_ch_vector}, as follows
\begin{equation}
\begin{split}
 \mathbf{H}&^{(m,n)}_{k,j} = \sum_{p =1}^{P}\boldsymbol{\eta}_{p,j} \boldsymbol{\eta}_{k,p} \\
 & = \sum_{p =1}^{P} \varepsilon_{p}e^{j 2 \pi (n T_0 \nu_{p} - m \Delta f \tau_{p})} \mathbf{b}(\theta_{p,j}, \gamma_{p,j}) \mathbf{a}^H(\phi_{k,p}, r_{k,p})
\end{split}
\end{equation}
where $v_p = v_{k,p} + v_{p,j}$ is the bistatic Doppler shift related to $p$th scatterer, whose expression will be given in Section~\ref{sec:sysmodelRangeDoppler}.
Thus, the $N^\mathrm{rf}_\mathrm{rx} \times 1$ vector of the received complex modulation symbols for a specific block $b$ is given by
\begin{align}
    \yv&_{b}[m,n] \nonumber\\ 
    & =\sum_{p=1}^{P} \varepsilon_{p} \Um_{b}^{H}\mathbf{b}(\theta_{p,j},\gamma_{p,j}) \mathbf{a}^H(\phi_{k,p},r_{k,p}) \nonumber \\ &  \fv_l~ x_{k}[m, n] e^{j2\pi(n\Tzero\nu_{p} - m\Delta f\tau_{p})}  + \wv[m, n].
 \label{eq:rx_sampled}
\end{align}
where $\wv \sim \mathcal{CN}(\mathbf{0},\sigma_\mathrm{w}^2 \mathbf{I}_{N^\mathrm{rf}_\mathrm{rx}})$ is the complex Gaussian noise.\\
%%%%%%%%%%%%%%%%%%%%%%%%%%%%%%%%%%%%%%%%%%%%%%%%%%%
By stacking the $M \times N$ \ac{OFDM} symbol grid into a $MN \times 1$ vector $\underline{\xv}_b$, where the underline symbol indicates blocked quantities, and defining $\Tm(\tau, \nu) \in \CC^{NM\times NM}$ as
\begin{multline}
    \Tm(\tau, \nu) = 
    {\rm diag}([1, \dots, e^{j2\pi n\Tzero\nu}, \dots, e^{j2\pi (N-1)\Tzero\nu}]^\trasp \\
    \otimes [1, \dots, e^{-j2\pi m\Delta f\tau}, \dots, e^{-j2\pi (M-1)\Delta f \tau}]^\trasp).
\end{multline}
% $\Tm = \text{diag}(e^{j2\pi(nT_0\nu_{p} - m\Df\tau_p)})$, 
the effective channel matrix of dimension $N^\mathrm{rf}_\mathrm{rx} M N \times M N$, defined for a single scatterer $p$ is given by
\begin{align} \label{eq:G_mat_full}
\Gm_b &(\tau_p,\nu_p, \theta_{p,j}, \phi_{k,p}) \eqdef \\ \nonumber & \Tm \otimes \left(\Um_{b}^{H}\mathbf{b}(\theta_{p,j},\gamma_{p,j}) \mathbf{a}^H(\phi_{k,p},r_{k,p}) \fv_l \right) 
\end{align}
The received signal $\underline{\yv}_b \in \mathbb{C}^{N^{\mathrm{rf}}_\mathrm{rx} M N \times 1}$ can then be written as

\begin{equation}
    \underline{\yv}_b = \
  \sum_{p=1}^{P} \varepsilon_{p}\Gm_b(\tau_p,\nu_p, \theta_{p,j}, \phi_{k,p}) \underline{\xv}_b + \wv_b .
   \label{eq:rx_blocked}
\end{equation}
%Considering $B$ blocks, the total received signal of dimension  $N^\mathrm{rf}_{rx} M N B \times 1$, is defined as $\yv = [\yv_1^T, \yv_2^T, \dots, \yv_B^T]^T$. 
For the derivation of the \ac{ML} target parameter estimation in Section~\ref{MLE_DET}, we consider a single \ac{Tx} and \ac{Rx} pair and drop their respective indices, $(.)_{k,j}$, to avoid excessive notation clutter. Needless to say, the same formulation holds for other pairs. 

% The analysis in this paper will be performed considering $B=1$, without loss of generality.

%%%%%%%%%%%%%%%%%%%%%%%%%%%%%%%%%%%%%%%%%%%%%%%%%%%%%%%%%%%%%%%%%%%%%%%%%%%%%%%%
\section{Parameter Estimation and Detection} \label{MLE_DET}
% \subsection{Maximum likelihood parameter estimation}\label{sec:Joint-Detection-Param-Estimation}
Since we assume no a priori knowledge of the target's initial position in regard to the near/far-field, we introduce a \textit{two-stage} \ac{ML} parameter estimation framework. In the initial stage, radial segments of the beam space are covered using \ac{FF} beamforming, and the target parameters are estimated based on bi-static \ac{FF} assumptions at the \ac{Rx} units. This scheme leads to a twofold advantage; 1) As depicted in Fig.~\ref{fig:codebook_scheme} covering the beam space with beamforming vectors reduces the latency resulting from codeword selection since the \ac{FF} codebook is parameterized only in angle, whereas the \ac{NF} codebook is two-dimensional. 2) As will be shown in this section, the parameter search space of the \ac{FF} model is significantly smaller than that of the \ac{NF} model, thereby reducing the parameter estimation complexity. 

For the second stage, based on the estimates obtained from the first stage,  if the target is determined to be in the \ac{NF} region of either the \ac{Tx} or \ac{Rx} units, a second estimation stage is performed. Depending on the target being located in the \ac{NF} of the \ac{Tx} or \ac{Rx}, the second stage will differ, as will be outlined in the following. 

Important to note that when the \ac{NF} cases occur in the initial stage, the model mismatch between the \ac{NF} and \ac{FF} array steering vectors presented in Section~\ref{subsection:PhysicalModel} leads to poorly localized estimates. However, these approximate and inaccurate estimates are used to define a \textit{region of interest} for further processing based on the correct (i.e. \ac{NF}) model.   
%%%%%%%%%%%%%%%%%%%%%%%%%%%%%%%%%%%%%%%%%%%%%%%%%%%%%%%%%%%%%%%%%%%%%%%%%%%%%%%
%%%%%%%%%%%%%%%%%%%%%%%%%%%%%%%%%%%%%%%%%%%%%%%%%%%%%%%%%%%%%%%%%%%%%%%%%%%%%%%%
%\subsection{Detection and maximum likelihood estimation} \label{detection}
\subsection{Maximum likelihood parameter estimation}\label{sec:Joint-Detection-Param-Estimation}
Considering that the delay parameter can be written as $\tau= (r+\gamma)/c$~, and therefore parameterizing $\Tm(\tau,\nu) \rightarrow \Tm(r, \gamma,\nu)$, we denote the true value of target parameters as $\mathring{\thetav} = \{\mathring{\varepsilon}_p, \mathring{\nu}_p, \mathring{r}_p, \mathring{\gamma}_p, \mathring{\theta}_p,\mathring{\phi}_p\}_{p=0}^{P-1}$. 
%and using $\thetav = \{h_p, \nu_p, \tau_p, \phi_p\}$ to express the arguments of the likelihood function, (I think this is not used...)
%As said before, assuming that $\mathring{\thetav}$ is constant over $B$ OTFS blocks, target detection and parameter estimation are performed from the $B$-block observation.  
% The received signal expression is written in a compact form by blocking the $NM$ Doppler-delay signal components into $NM \times 1$ vectors, where the underline symbol indicates blocked quantities. 
% We define $\Tm(\tau, \nu) \in \CC^{NM\times NM}$ as:
% \begin{multline}
%     \Tm(\tau, \nu) = 
%     {\rm diag}([1, \dots, e^{j2\pi nT\nu}, \dots, e^{j2\pi (N-1)T\nu}]^\trasp \\
%     \otimes [1, \dots, e^{-j2\pi m\Delta f\tau}, \dots, e^{-j2\pi (M-1)\Delta f \tau}]^\trasp).
% \end{multline}
For each $b \in[B]$, the  effective channel matrix of dimension $\Nrf NM\times NM$ associated with scattering point $p$ is given as ${\Gm}_{b}(\mathring{\nu}_{p},\mathring{r}_{p}, \mathring{\gamma}_{p},\mathring{\theta}_{p}, \mathring{\phi}_{p})$ in \eqref{eq:G_mat_full}. The received signal then takes the form 
\begin{align}
    \underline{\yv}_{b}= \sum_{p=0}^{P-1} \mathring{\varepsilon}_{p} \Gm_b(\mathring{\nu}_{p},\mathring{r}_{p}, \mathring{\gamma}_{p},\mathring{\theta}_{p}, \mathring{\phi}_{p}) \underline{\xv}_b + \wv_b .
   \label{eq:rx_blocked_true}
\end{align}
Given knowledge of the number of scattering points $P$, the \ac{ML} estimate of the set $\mathring{\thetav}$ can be obtained by solving \eqref{eq:full_ML},
% \begin{align}\label{eq:full_ML},
%     &\thetav_{\rm ML} = \\
%     & \underset{\{\epsilon_{p}, \nu_{p}, r_p, \gamma_{p},\theta_p, \phi_{p}\}_{p=0}^{P-1}\in\Gamma}{\arg\min}\sum_{b=0}^{B-1}\left\| \underline{\yv}_{b} - \sum_{p=0}^{P-1} \epsilon_{p} \Gm_b(\nu_{p}, r_p,\gamma_{p}, \theta_p,\phi_{p}) \underline{\zetav}_b \right\|_2^2 \nonumber,
% \end{align}
\newcounter{TempEqCnt}
\setcounter{TempEqCnt}{\value{equation}}
\setcounter{equation}{26}
\begin{figure*}[ht]
% \begin{multline}
\begin{align}
    \thetav_{\rm ML} = 
    \underset{\{\varepsilon_{p}, \nu_{p}, r_p, \gamma_{p},\theta_p, \phi_{p}\}_{p=0}^{P-1}\in\Gamma}{\arg\min}\sum_{b=0}^{B-1}\left\| \underline{\yv}_{b} - \sum_{p=0}^{P-1} \varepsilon_{p} \Gm_b(\nu_{p}, r_p,\gamma_{p}, \theta_p,\phi_{p}) \underline{\xv}_b \right\|_2^2 , ~\label{eq:full_ML}
\end{align}
% \end{multline}
\hrulefill
\end{figure*}
where the space is $\Gamma\eqdef\CC^P\times\RR^{5P}$. 

Solving \eqref{eq:full_ML} requires knowledge of the number of scattering points $P$, which can be formulated as a model order estimation problem. Since the \textit{micro-scatterers} of extended targets are often indistinguishable, the estimation of the model order is an unattainable task due to the problem being intrinsically ill-posed \cite{Rife}. In addition, the parameter space of the brute force \ac{ML} solution in \eqref{eq:full_ML} requires prohibitively large computations. Therefore, we resort to an approximate method that evaluates a hypothesis test on a set of $(\nu_p, r_p, \gamma_p,\theta_p, \phi_p)$ tuples belonging to a  grid $\Theta$. This approach involves a detection and estimation step for each parameter tuple. Specifically, we formulate the target detection as a standard Neyman-Pearson hypothesis testing problem \cite{VPoor}, for which the solution that maximizes the detection probability subject to a bound on the false-alarm probability is given by the Likelihood Ratio Test with hypotheses $\Hc_0$ and $\Hc_1$ corresponding to the absence or presence of a target (see \cite{Dehk_TWC} for details). 
To formulate this problem, the log-likelihood ratio for the binary hypothesis testing problem given by 
\begin{align}
    \ell(\varepsilon,\nu, r, \gamma,\theta, \phi) & = \log \frac{\exp\left ( - \frac{1}{\sigma_w^2} \sum_{b=1}^B \left \| \underline{\yv}_b - \varepsilon \underline{\Gm}_b \underline{\xv}_b \right \|_2^2 \right )}{\exp \left ( - \frac{1}{\sigma_w^2} \sum_{b=1}^B \| \underline{\yv}_b \|_2^2 \right )} \nonumber \\
    % & = 2 \Re\left \{ \left ( \sum_{b=1}^B \underline{\yv}_b^\herm \underline{\Gm}_b \underline{\xv}_b \right ) h_p \right \} - |h_p|^2 \sum_{b=1}^B \| \underline{\Gm}_b \underline{\xv}_b \|^2 
    \label{log-likelihood2}
\end{align}
is compared against a threshold for every point in $\Theta$, resulting in the generalized likelihood ratio test
\begin{equation} \label{GLRT}
    % \ell(\nu_p, r_p, \gamma_p,\theta_p, \phi_p)  \underset{\Hc_0}{\overset{\Hc_1}{\gtrless}} T_r~,\quad (\nu_p, r_p, \gamma_p,\theta_p, \phi_p) \in \Theta
        \ell(\nu, r, \gamma,\theta, \phi)  \underset{\Hc_0}{\overset{\Hc_1}{\gtrless}} T_r~,\quad (\nu, r, \gamma,\theta, \phi) \in \Theta
\end{equation}
where the threshold $T_r$ is chosen at each grid point by using the \ac{OS-CFAR} technique described in \cite{MultiDimCFAR}.
For the binary hypothesis test above, the (generalized) log-likelihood ratio in \eqref{log-likelihood2} coincides with the likelihood function \eqref{eq:RxU_ml_estimate_opt}, derived in Appendix \ref{chap:app_mle} and is given by
\begin{align} \label{eq:RxU_ml_metric}
    \ell(\nu, r, \gamma,\theta, \phi) = \frac{\left\lvert \fv^\herm \av \bv^\herm \xiv_{(B)}(r, \gamma,\nu) \right\rvert^2}{\fv^\herm \av \bv^\herm \bar{\Um}_{(B)} \bv\av^\herm \fv}.
\end{align}
Note that the \ac{ML} function in \eqref{eq:RxU_ml_estimate_opt} is valid considering a single scattering point. When dealing with multi-target scenarios that are well-separated (w.r.t. system resolution limits), recovery of the numerous scatterers requires an \ac{SIC} technique, where the contribution of the estimated point is removed from the signal, and the metric is re-evaluated in an iterative manner until some stopping criteria are met. In extended-target scenarios where the micro-scatterers are often very closely spaced, the mutual influence caused by the ``sidelobes’’ of the likelihood function for adjacent micro-scatterers renders dependency on the hypothesis testing for adjacent points. Therefore \ac{SIC} techniques can be ineffective. However, as it will be shown in Section~\ref{simulation_results}, the simple grid-based estimation and thresholding (detection) approach proposed above leads to satisfactory performance.
Since we assume that the target location within the \ac{NF} or \ac{FF} of the arrays is initially unknown, to tackle the large search space of the refined grids over $\Theta$ to obtain highly accurate estimates, we propose the two-stage \ac{ML} estimation method that follows. 

\begin{remark}
   The term target ``detection'' in this section differs from that implying the detection of a target such as a vehicle as a potential communication \ac{UE} described in Section~\ref{sec:phy-model} which can be achieved via communication. In the grided \ac{ML} search above, the detection corresponds to distinguishing the response from a point scatterer (or multiple point scatterers that violate the system resolution limits and therefore show as a point response) from noise in the radar image, produced from evaluating \eqref{eq:RxU_ml_metric} over the set of tuples on the defined grid.   
\end{remark}
%{\RED maybe remove this remark or move to another place in text}
%\begin{remark}
%    Note that in classical \ac{FF} models, when dealing with extended targets, it is probable that very closely- spaced scattering points (in spatial domain i.e. angle and range) cannot be distinguished due to system resolution limits. However, this assumption does not necessarily hold using the full array response in \eqref{eq:array_response}, since the time delay parameter can be extracted from the array response.
%\end{remark}
%%%%%%%%%%%%%%%%%%%%%%%%%%%%%%%%%%%%%%%%%%%%%%%%%%%%%%%%%%%%%%%%%%%%%%%%%%%%%%%%

\noindent \textbf{Stage 1: Far-field beamforming and bi-static estimation:}
In the first stage, we assume no knowledge of the target position. The appropriate \ac{FF} beamforming codeword is selected at the \ac{Tx}, and based on the assumption that the extended target is located in the \ac{FF} of both the \ac{Tx} and \ac{Rx} units, the array manifolds at the \ac{Tx} and \ac{Rx} units in \eqref{eq:G_mat_full}, which are only functions of angular parameters, i.e. ($\av(\phi,r)\rightarrow \av(\phi)$  and $\bv(\theta,\gamma)\rightarrow \bv(\theta)$ ) are considered for parameter estimation. Additionally, since the delay cannot be estimated from the array manifold in this model, it has to be estimated based on the observed total time of flight $\tau = (r+\gamma) /c$ from the sub-carrier dimension of the \ac{OFDM} frames, i.e. the phase observed on the $m\Df\tau$ component of the exponent in \eqref{eq:rx_sampled}. The bi-static Doppler projection seen by each of the \ac{Rx} units, given in \eqref{eq:R2}, can be estimated over the \ac{OFDM} symbol dimension, i.e. the phase observed on the $nT_0\nu$ component of the exponent in \eqref{eq:rx_sampled}  \cite{Pucci_Bistatic}. As such, the effective channel matrix $\Gm_b(\nu_{p}, r_p,\gamma_{p}, \theta_p,\phi_{p}) $ in \eqref{eq:G_mat_full} takes the form 
\begin{align} \label{eq:G_mat_ff_model}
\breve{\Gm}_b (\nu_{p},\tau_{p}, \theta_{p}) \eqdef \Tm \otimes \left(\Um_{b}^{H}\mathbf{b}(\theta_{p}) \mathbf{a}^H(\phi_{p}) \fv (\hat{\phi}_{p})\right). 
\end{align}
Note that, from the bi-static \ac{Rx}'s perspective, the channel response from \ac{Tx} to the target is a constant which can be absorbed in the channel gain coefficient, i.e. :
\begin{align}
    g_{p} \eqdef \av^H(\phi_p) \fv(\hat{\phi}_{p}) ~,~ h_p \eqdef g_{p} \varepsilon_p \nonumber
\end{align}
and therefore, $\breve{\Gm}_b$ is not a function of the \ac{AoD}, $\phi$. 
Then, defining the true value of parameters as $\mathring{\thetav} = \{\mathring{h}_p, \mathring{\nu}_p, \mathring{\tau}_p, \mathring{\theta}_p\}_{p=0}^{P-1}$ 
%and using $\thetav = \{h_p, \nu_p, \tau_p, \phi_p\}$ to express the arguments of the likelihood function, (I think this is not used...)
%As said before, assuming that $\mathring{\thetav}$ is constant over $B$ OTFS blocks, target detection and parameter estimation are performed from the $B$-block observation.  
the received signal then takes the form 
\begin{align}
    \underline{\yv_{b}} = \sum_{p=0}^{P-1} \mathring{h}_{p} \breve{\Gm}_b(\mathring{\nu}_{p}, \mathring{\tau}_{p}, \mathring{\theta}_{p}) \underline{\xv}_b + \wv_b .
   \label{eq:rx_blocked_ff}
\end{align}
Where for an extended target with  $P$ scattering points, the \ac{ML} estimate of the set $\mathring{\thetav}$ involves a search 
% Where for an extended target with  $P$ scattering points, the \ac{ML} estimate of the set $\mathring{\thetav}$ can be obtained by solving
% \begin{align}\label{eq:full_ML2}
%     &\thetav_{\rm ML}^{\rm FF}  = \\
%     &\underset{\{h_{p}, \nu_{p}, \tau_{p}, \theta_{p}\}_{p=0}^{P-1}\in\Gamma_{\rm FF}}{\arg\min}\sum_{b=0}^{B-1}\left\| \underline{\yv}_{b} - \sum_{p=0}^{P-1} h_{p} \breve{\Gm}_b(\nu_{p}, \tau_{p}, \theta_{p}) \underline{\xv}_b \right\|_2^2, \nonumber
% \end{align}
in a $ \Gamma_{\rm FF} \eqdef\CC^P\times\RR^{3P}$ space. Similar to the derivation of the likelihood function in the previous section, the likelihood function of the \ac{FF} model is obtained, which is omitted here for brevity. Note that in this case, the maximization of the log-likelihood function is performed with respect to $h_p$ for fixed $(\nu, \tau, \theta)$ to obtain:
% \begin{align} \label{eq:RxU_ml_estimate_ff}
%    \ell (\hat{\nu}_p, \hat{\tau}_p,\hat{\phi}_p) = \underset{\nu,\tau,\phi}{\rm argmax} \frac{\left\lvert \bv^\herm \xiv_{(B)}(r, \gamma,\nu) \right\rvert^2}{\bv^\herm \bar{\Um}_{(B)} \bv},
% \end{align}
\begin{align} \label{eq:RxU_ml_estimate_ff}
   \ell (\nu, \tau,\theta) = \frac{\left\lvert \bv^\herm \xiv_{(B)}(\tau,\nu) \right\rvert^2}{\bv^\herm \bar{\Um}_{(B)} \bv},
\end{align}
By defining a suitably refined search grid on $\Theta_{\rm FF}\eqdef\RR^{3P}$, and evaluating \eqref{eq:RxU_ml_estimate_ff} for every tuple $(\nu, \tau, \theta)~\in \Theta_{\rm FF}$ and performing the estimation and detection (thresholding) step according to Section~\ref{sec:Joint-Detection-Param-Estimation}, the estimates $(\hat{\nu}, \hat{\tau}, \hat{\theta})$ of the scattering points can be obtained. To convert these values to the angle and range of the target in the global coordinates, we use the bi-static conversion principles in the next section. 

%%%%%%%%%%%%%%%%%%%%%%%%%%%%%%%%%%
\subsection{Bi-static range and Doppler shift}\label{sec:sysmodelRangeDoppler}
%%%%%%%%%%%%%%%%%%%%%%%%%%%%%%%%%%
In a bi-static configuration, the propagation time $\tau_p$ of the signal scattered by a scatterer $p$ is related to the distance between the \ac{Tx} and the scatterer, $r_{\mathrm{tx},p}$, and that between the scatterer and the Rx, $\gamma_{p,\mathrm{rx}}$, via the bi-static range, $R_\mathrm{bis} = r_{\mathrm{tx},p} + \gamma_{p,\mathrm{rx}} = \tau_p \cdot c$ \cite{Pucci_Bistatic}.
After estimating $R_{\mathrm{bis}}$ via $\tau_p$, the scatterer can be located on an ellipse with a major axis equal to $R_\mathrm{bis}$ and foci at \ac{Tx} and \ac{Rx} positions, as depicted in the inset of Fig.~\ref{fig:scene_topology}.
The \ac{Tx}, \ac{Rx}, and scatterer form a triangle with base $\Delta$ (with $\Delta$ the distance between \ac{Tx} and \ac{Rx}) called the baseline; the angle $\beta$ of the opposite vertex is named the bi-static angle.

If the \ac{AoA} $\theta_{p,\mathrm{rx}}$ of the reflected echo at the \ac{Rx} can be estimated, it is possible to determine the distance $r$ as \cite{Willis05}
\begin{equation} \label{eq:R2}
    \gamma_{p,\mathrm{rx}} = \frac{R_\mathrm{bis}^2-\Delta^2}{2(R_\mathrm{bis}+\Delta\sin{(\theta_{p,\mathrm{rx}}-\pi/2)})}
\end{equation}
and then the scatterer position 
\begin{equation}
    \mathbf{p}_p = (x_\mathrm{rx}-\gamma_{p,\mathrm{rx}}\cos{\theta_{p,\mathrm{rx}}},y_\mathrm{rx}+\gamma_{p,\mathrm{rx}}\sin{\theta_{p,\mathrm{rx}}}).
\end{equation}

In addition to the scatterer location, the bi-static velocity of the scatterer can be inferred from the bi-static Doppler shift. The latter is proportional to the rate of change of $R_\mathrm{bis}$. When \ac{Tx} and \ac{Rx} are stationary, and the scatterer is moving with velocity $\mathbf v_p$, the Doppler shift can be obtained as \cite{Willis05}
\begin{align}
    \nu_p & = \nu_{\mathrm{tx},p} + \nu_{p,\mathrm{rx}}\nonumber \\
    & = \frac{1}{\lambda_\mathrm{c}}\frac{d}{dt}\left[r_{\mathrm{tx},p}(t)+\gamma_{p,\mathrm{rx}}(t)\right] = \frac{2v_p}{\lambda_\mathrm{c}}\cos{\delta} \cos{(\beta/2)} 
    \label{f_D}
\end{align}
where $\delta$ is the angle between the direction of the velocity and the bi-static bisector, and $v_p=|\mathbf v_p |$.
While $\beta$ can be easily determined by knowing $\Delta$, $r_{\mathrm{tx},p}$, $\gamma_{p,\mathrm{rx}}$, and $\theta_{p,\mathrm{rx}}$, the angle $\delta$ is unknown so only the bi-static velocity, $v_\mathrm{bis} = |\mathbf{v}_\mathrm{bis}| = v_p \cos{\delta}$, can be estimated by the system.

\noindent \textbf{Stage 2: Near-field estimation}

The second stage is only carried out if the target is determined to be in the NF of the \ac{Tx} or \ac{Rx} units, based on the estimates obtained by evaluating the \ac{FF} model in \eqref{eq:RxU_ml_estimate_ff}. Below we describe the second stage for each scenario.

\noindent \textbf{Near-field of \ac{Rx}: Parameter estimation with reduced search space:} Assume the target is determined to lie in the \ac{NF} of an \ac{Rx} array. This means the \ac{ML} metric in \eqref{eq:RxU_ml_metric} needs to be evaluated on a fine-grained grid defined over $\Theta$ to meet high accuracy localization requirements. It is clear that evaluating a 5-D search grid is computationally heavy. Therefore, we define a suitably refined Cartesian grid over the \ac{RoI}, indicated by the approximate estimates from the first stage. Subsequently, the coordinates of each cell in the grid are translated to the equivalent $(r_p, \gamma_p,\theta_p, \phi_p)$ and the \ac{ML} metric in \eqref{eq:RxU_ml_estimate_opt} is evaluated.
%%%%%%%%%%%%%%%%%%%%%%%%%%%%%%%%%%%%%%%%%%%%%%%%%%%%%%%%%%%%%%%%%%%%%%%%%%%%%%%%
% \begin{figure}
% \centering
% \resizebox{0.75\linewidth}{!}{\input{Plots_Figures/NF_est_2d}} %resize the content to a fixed width
% \caption{ML estimates.}
% \label{fig:nf_est_ml}
% \end{figure}

% \begin{figure*}[ht]
% \centering
% \scalebox{0.8}{\includegraphics{Plots_Figures/NF_MLE_GT.pdf}}
% \caption{\small The figure shows an extended target located in the \ac{NF} of Rx$_1$. After performing an \ac{ML} estimation based on \ac{FF}-bistatic assumptions, the estimated scattering points from Rx$_1$ indicate the target to be in the \ac{NF} of this Rx unit. Then, a fine-grained search grid is defined in the indicated \ac{RoI}, where the \ac{ML} metric in \eqref{eq:RxU_ml_estimate_opt} is evaluated as shown on the left. }
% \label{fig:NF_MLE} 
% \end{figure*}

% \begin{figure*}[ht]
% \centering
% \scalebox{0.8}{\includegraphics{Plots_Figures/NF_MLE_GT2.pdf}}
% \caption{\small The figure shows an extended target located in the \ac{NF} of Rx$_1$. After performing an \ac{ML} estimation based on \ac{FF}-bistatic assumptions, the estimated scattering points from Rx$_1$ indicate the target to be in the \ac{NF} of this Rx unit. Then, a fine-grained search grid is defined in the indicated \ac{RoI}, where the \ac{ML} metric in \eqref{eq:RxU_ml_estimate_opt} is evaluated as shown on the left. }
% \label{fig:NF_MLE} 
% \end{figure*}

%%%%%%%%%%%%%%%%%%%%%%%%%%%%%%%%%%%%%%%%%%%%%%%%%%%%%%%%%%%%%%%%%%%%%%%%%%%%%%%%
\subsubsection*{Discussion} Note that the transition region between \ac{NF} and \ac{FF} does not have a hard cut-off, in the sense that \ac{FF} beams already start to form after approximately $D_{\rm ff}/10$ \cite{Bjorn_Primer, Schober_2023} with some phase variations. Therefore, as the radial distance from the array approaches this distance, the array manifold tends increasingly toward the \ac{FF} model. This is important in that if a target is located in these transitional regions, it will not be critical to determine the \ac{NF} or \ac{FF} regime, and even the \ac{FF} model will result in acceptable performance for parameter estimation. This effect is also evident from numerical results provided in Section \ref{beamfocusing_results}, where the beamforming and beamfocusing schemes exhibit similar performance in these regions.       

\noindent \textbf{Near-field of \ac{Tx}: Beamfocusing and re-estimation}

If the target is determined to be in the \ac{NF} of an \ac{Tx} array, this means that the illumination beamformer at the \ac{Tx} cannot focus the beam in the intended location with full \ac{BF} gain. In this case, a codeword from a custom-designed beamfocusing codebook is selected to focus the energy in the area estimated by the first stage. These codewords are designed (see Section~\ref{BeamFocusing} for details) to maintain a constant gain in an extended region (angle-range). This scheme provides a two-fold advantage. 1) Due to the increase in \ac{BF} gain (and therefore higher SNR) after selecting the appropriate codeword, a re-estimation of the target parameters at the \ac{Rx} leads to more accurate estimates. Note that in this case, the \acp{Rx} can use the \ac{FF} model due to the deployment topology (see Fig.~\ref{fig:scene_topology}). 2) In the possible case that the target is also a communication user, the communication performance can be significantly increased. Since the beamfocusing illuminates an extended region with a constant gain if the true \ac{UE} antenna location deviates slightly from the one estimated from the back-reflected signal, a good \ac{SNR} can still be maintained. We further remark that, if the region of interest for the second stage is larger than the area covered by the \ac{NF} codeword, multiple neighboring codewords can be used in time-division manner to cover the \ac{RoI} for re-estimation.      

%%%%%%%%%%%%%%%%%%%%%%%%%%%%%%%%%%%%%%%%%%%%%%%%%%%%%%%%%%%%%%%%%%%%%%%%%%%%%%%%
%%%%%%%%%%%%%%%%%%%%%%%%%%%%%%%%%%%%%%%%%%%%%%%%%%%%%%%%%%%%%%%%%%%%%%%%%%%%%%%%
\section{Design of the beamfocusing weights} \label{BeamFocusing} 
%%% Pseudocode Algorithm %%%
\begin{algorithm}[hbt!]
\caption{Algorithm for designing beamfocusing vectors}\label{alg:beamspot_design}
\begin{algorithmic}
%\Require $\epsilon$, \Bm \Comment{$\epsilon$ is solution tolerance}
%\Ensure $y = x^n$
\For{codeword g $\in \CBf$}
     \begin{enumerate}
            \item 	choose solution tolerance $\varrho$, Tikhonov regularizer $\epsilon_T$,  set $\cv_g = e^{j*\zerov_G}$ 
            \item 	initialize $\fv_g$ \Comment{see \textbf{Initialization}}
             \item  set $\cv_g$ to $\cv_g = e^{j\angle(\Am^H \fv_g)}$
             \item fix $\cv_g$, update new $\fv_g$ by obtaining the residual error from evaluating the unconstrained linear least-squares problem, $\| \bv_g-|\Am^H  \fv_g| \|_2$.
              % \item fix $\cv_g$, update new $\fv_g$ as solution to the unconstrained linear least-squares problem, $\| \bv_g-|\Am^H  \fv_g| \|_2$.
             \item repeat steps (2 - 4) until the decrease in objective function has diminished to within $\varrho$.
    \end{enumerate}
\EndFor

\vspace{3mm}
\State \textbf{Initialization}
\State $\yv' \gets \yv_g \odot\cv_g$ \Comment{$\yv_g$ is the mask corresponding to $\fv_g$}
\State $\fv_g \gets (\Am\Am^H + \epsilon_T\Id_{\Na})^{-1}\Am \ \yv'$
% \While{$N \neq 0$}
% \If{$N$ is even}
%     \State $X \gets X \times X$
%     \State $N \gets \frac{N}{2} $  \Comment{This is a comment}
% \ElsIf{$N$ is odd}
%     \State $y \gets y \times X$
%     \State $N \gets N - 1$
% \EndIf
% \EndWhile
\end{algorithmic}
\end{algorithm}

% %%%%%%%%%%%%%%%%%%%%%%%%%%%%%%%%%%%%%%%%%%%%%%%%%%%%%%%%%%%%%%%%%%%%%%%%%%%%%
\subsection{Related works}
With increasing carrier frequencies of wireless communication networks and the deployment of large arrays, the \ac{NF} region is significantly expanded. Thus \ac{FF}-based beamforming techniques such as \ac{DFT} codebooks can result in significant \ac{SNR} losses, affecting both the communication and sensing performance of the system. More recently, there have been numerous works dealing with \ac{NF} beamfocusing schemes, especially in the Reconfigurable Reflecting Surfaces domain, due to the large array sizes used therein \cite{bjornseen,ring_type2}. While a few works have investigated using beamfocusing weights obtained by conjugating the \ac{NF} array response, others have resorted to optimization-based methods to obtain suitable weights \cite{Wang_2023,NF_dardari}. The most significant drawback of those methods is that a very accurate estimate of the intended user coordinates (equivalently \textit{range} and \textit{angle}) is required (see \eqref{eq:array_response} for the array response). Even when ignoring the cost of obtaining such estimates, these techniques pose another significant challenge since the goal is to focus beams on the \ac{UE}'s antenna. In the very likely scenario of physically extended targets (e.g. motorbikes, bicycles, cars, etc.) the \ac{UE} antenna can be located anywhere on the object and the estimated reflection points do not necessarily correspond to the reflections from the antenna.
In other works \ac{NF} codebook-based techniques have been employed. Some of the more promising approaches are the \textit{Ring-type Codebook} designs \cite{ring_type2},  where a first layer phase distribution is calculated based on the Fresnel principle and is then superimposed with the
codeword selected of a (\ac{FF}) \ac{DFT} codebook in the second layer. As shown in \cite{ring_type2}, such design lead to significant spectral leakages in unwanted locations which beats the purpose of user-interference reduction via beamfocusing. From the sensing perspective, this leads to reflection from unintended objects that may be located in the undesired illuminated areas. 
Considering the issues mentioned above, we devise a codebook-based scheme inspired by flat-top beamforming techniques that provide an (almost-) constant gain over an extended angular span (see \cite{Dehk_TWC} and references therein for details). In the case of beamfocusing for the \ac{NF}, we present a method that aims to synthesize beamfocusing weights for array operation in \ac{NF} such that an extended area is illuminated with a relatively constant gain. 

As a last note, it should be pointed out that one could essentially view \ac{NF} beamfocusing optimization problem as a 2-dimensional filter design problem. These problems have previously extensively been studied in \cite{2d_svd, Lu_sdp}. A very similar approach is the design of 2-D spatial filters in the form of azimuth-elevation response (i.e. radiation pattern) of antenna arrays (see e.g. \cite{2d_pattern_syn}).

\subsection{Problem formulation}
Let $\fv$ be a beamfocusing vector of dimension $\Na$. The complex-valued (amplitude and phase) 
beam pattern radiated by the array at each sampling tuple $(\tilde{\phi}_i,\tilde{r}_j),~ i\in [G_{\theta}],~ j\in [G_{r}]$ of a discrete angular set $\{\tilde{\Omega}\},~ (|\tilde{\Omega}| = G_{\theta})$ and range set $\{\tilde{\Gamma}\},~ (|\tilde{\Gamma}| = G_{r})$ can be calculated as the inner product of the vector $\fv$ and the array response vector $\av(\phi,r)$ at the given grid angle-range tuple i.e., $\av^H(\tilde{\phi}_i,\tilde{r}_j)\fv$. 

The design problem of interest is to find $\fv$ to approach a desired 
radiation pattern $\bar{\bv} \in \RR_{+}^{G}$. The entries of $\bar{\bv} = [\bar{b}_1,...,\bar{b}_{G}]$ are magnitudes of the radiation pattern at each of the $G=G_{\theta}G_r$ discrete tuple points. In particular, we fix $\bar{\bv}$ to have a constant level in a pre-determined angle-range zone (i.e. spot) in the \ac{NF} of the array and such that the values
corresponding to the rejection directions (sidelobes) are below a certain threshold with respect to the maximum (center beam). 
By letting $\Am =[\av(\tilde{\phi}_1,\tilde{r}_1), \dots, \av(\tilde{\phi}_{G_{\theta}},\tilde{r}_{G_{r}})] \in \RR^{\Na\times G}$, this problem can be formulated as a magnitude least-squares problem which belongs to the class of problems addressed by \cite{MLS_2, Kassakian}. 
\begin{align}
    \min_{\fv} & \quad \| \Am^H \fv-\bar{\bv}\|_2^2 \nonumber\\ \label{eq: bp_optim_1}
    \text{s.t.} & \quad  \fv^H \Am \Am^H\fv = 1
\end{align}
%\begin{equation} \label{eq: bp_optim_1}
%\begin{split}
%\mathcal{P}(1):\quad &\underset{\fv}{\min} %\quad\left\{\sum^{N_a}_{j=1}(\vert \matr{A}^{T}_{:,j}\fv\vert -\bar{\bv}_{j})^{2}\right\} \\
%& \text{s.t.}\quad\quad \fv^{H}\Am^{H}\Am\fv=1 %p
%\end{split}
%\end {equation}  
where the constraint in \eqref{eq: bp_optim_1} imposes unit transmit power. 
Depending on the operating scenario, a beam pattern can focus the transmitted energy on a certain given angle-range sector (i.e., \ac{FoV} equal to $\Omega \times \Gamma$). In order to define our design in a flexible manner, the \ac{FoV} is divided into multiple sectors as depicted in Fig.~\ref{fig:beam_spots_CB}, each sector determining the illumination area of a codeword (i.e. the span of the beamfocusing \textit{spots}). These codewords are gathered in a codebook $\CBf$. Note that the problem in \eqref{eq: bp_optim_1} aims to approximate the phase and magnitude of the reference mask, which gives a more constrained, but less localized solution, as required by our particular application. To relax this, we consider only the magnitude response in the optimization problem:
\begin{align}
    \min_{\fv} & \quad \| ~|\Am^H \fv| -\bar{\bv}\|_2^2 \nonumber\\ \label{eq: bp_optim_2}
    \text{s.t.} & \quad  \fv^H \Am \Am^H\fv = 1
\end{align}

The above problem is not convex, and therefore by performing a semidefinite relaxation, acceptable solutions can be obtained. Important to note that, in the case of one–dimensional filters and uniformly spaced
linear arrays, one–dimensional spectral factorizations of magnitude responses are guaranteed to exist, and therefore the problem is significantly less complex than a multidimensional case, such as in this problem \cite{Kassakian}. By defining $\Bm = \text{diag}(\bar{\bv})$ and restructuring the objective in \eqref{eq: bp_optim_2} as:
\begin{align}
    \min_{\fv,\cv} & \quad \| ~\Am^H \fv -\Bm \cv\|_2^2 \nonumber\\ \label{eq: bp_optim_3}
    \text{s.t.} & \quad |\cv| = \textbf{1}
\end{align}
It is possible to minimize firstly over $\fv$ and then minimize over $\cv$, leading to:
\begin{align}
    \min_{|\cv| = 1}~(\min_{\fv} &~\| ~\Am^H \fv -\Bm \cv\|_2^2) \label{eq: bp_optim_4}
\end{align}
Where, with the assumption that $\Am$ has full rank, the inner minimization problem has an analytic solution w.r.t. $\cv$, i.e. $(\Am \Am^H)^{-1}\Am\Bm\cv$. Define $\Wm = \Am^H(\Am \Am^H)^{-1}\Am \Bm - \Bm$, then the problem can be formulated as below:
\begin{align}
    \min_{\cv} & ~ ~\cv^H \Wm^H \Wm \cv\quad  \nonumber\\ \label{eq: bp_optim_5}
    \text{s.t.} & \quad |\cv| = \textbf{1}
\end{align}
Note that, due to the fine-grained grid which is used for the sampling points (i.e. $G_r, G_{\theta}$) of the array manifold matrix $\Am$, this matrix is often not full rank (i.e. rank $(\Am)$ $< \#\text{cols}(\Am)$~), and therefore the solution in \eqref{eq: bp_optim_4} does not hold.  To overcome this one can use a \textit{Tikhonov} regularization with parameter $\epsilon_T~\in \RR_{+}$ and re-formulate the problem as:

\begin{align}
    \min_{\fv,\cv} & \quad \| ~\Am^H \fv -\Bm\cv\|_2^2 + \epsilon_T \| \fv\|_2^2 \ \label{eq: bp_optim_tikh}
\end{align}

Then, equivalently by defining $\Um = \Am^H(\Am\Am^H+\epsilon_T\Id_{\Na})^{-1}\Am- \Id$ and  $\Wm=(\Um\Bm)^{H}(\Um\Bm)$ and 
\begin{align}
    % \tilde{\Wm} = [\Re{\Wm}, -\Im{\Wm}; \Im{\Wm}, \Re{\Wm}] \nonumber
    \tilde{\Wm} =
    \begin{bmatrix}
\Re{(\Wm)} & -\Im{(\Wm)} \\
\Im{(\Wm)} & \Re{(\Wm)},
\end{bmatrix}
\end{align}
and defining $\Vm=\epsilon((\Am\Am^H+\epsilon_T\Id_{\Na})^{-1}\Am\Bm)^H(\Am\Am^H+\epsilon_T\Id_{\Na})^{-1}\Am\Bm))$, and $\tilde{\Vm}$ in similar fashion as $ \tilde{\Wm}$, the following problem should be solved:
\begin{align}
    \min_{\Cm} & ~~\text{trace}(\Cm \tilde{\Wm}) +  \text{trace}( \Cm \tilde{\Vm} ) \nonumber\\ 
    \text{s.t.} & \quad  \Cm_{ii} + \Cm_{jj}= 1 \nonumber\\ 
     & \quad  \Cm ~~\succcurlyeq 0
     \label{eq:bp_optim_tikhonov2}
\end{align}
Then, the beamforming vector can be found as $\fv=(\Am\Am^{H} + \epsilon_T\Id_{\Na})^{-1}\Am\Bm\Cm$. Above, $\Cm \in \RR^{2G\times 2G}$ is diagonalized from the real and imaginary part of $\cv$ as $\Cm = \text{diag}(\cv_r\cv_r^T,\cv_i\cv_i^T)$.

\begin{figure*}[ht]
\centering
\scalebox{0.35}{\begin{tikzpicture}[auto,rotate=0,transform shape]
\begin{axis}[%
width=4.4in,
height=4.4in,
at={(0in,0in)},
scale only axis,
axis on top,
xmin=0.5,
xmax=11.9,
xlabel style={font=\color{white!15!black},font=\Large},
xlabel={Range [m]},
xtick={1,3,...,11},
y dir=reverse,
ymin=-20,
ymax=20,
ytick={-18,-12,...,18},
ylabel style={font=\color{white!15!black},font=\Large},
ylabel={Angle [deg]},
axis background/.style={fill=white}
]
\addplot [forget plot] graphics [xmin=0.5, xmax=11.9, ymin=-20, ymax=20] {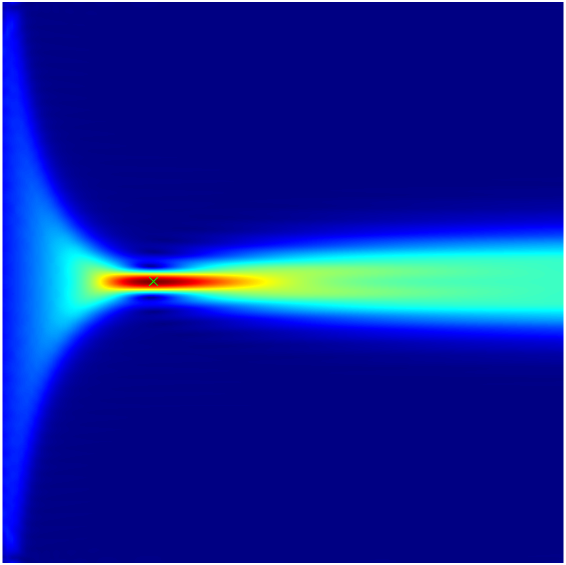};
\end{axis}
\end{tikzpicture}%

\hfill

\begin{tikzpicture}[auto,rotate=0,transform shape]

\begin{axis}[%
width=4.4in,
height=4.4in,
at={(0in,0in)},
scale only axis,
axis on top,
xmin=0.5,
xmax=11.9,
xlabel style={font=\color{white!15!black},font=\Large},
xlabel={Range [m]},
xtick={1,3,...,11},
y dir=reverse,
ymin=-20,
ymax=20,
ytick={-18,-12,...,18},
ylabel style={font=\color{white!15!black},font=\Large},
ylabel={Angle [deg]},
axis background/.style={fill=white}
]
\addplot [forget plot] graphics [xmin=0.5, xmax=11.9, ymin=-20, ymax=20] {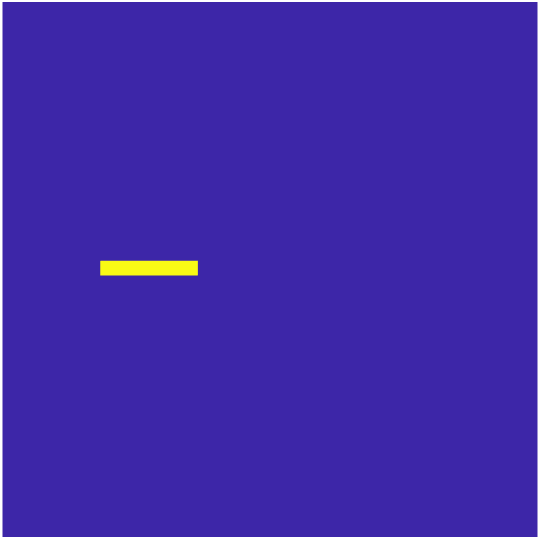};
\end{axis}
\end{tikzpicture}%
%%%%%%%%%%%%%%%%%%%%%%%%%%%%%%%%%%%%%%%%%%%%%%%%%%%%%%%%%%%%%%%%%%%%%%%%%%%%%%%%%%%%%%%%%%%%%%%%%%

\begin{tikzpicture}[auto,rotate=0,transform shape]
\begin{axis}[%
width=4.4in,
height=4.4in,
at={(0in,0in)},
scale only axis,
axis on top,
xmin=0.5,
xmax=11.9,
xlabel style={font=\color{white!15!black},font=\Large},
xlabel={Range [m]},
xtick={1,3,...,11},
y dir=reverse,
ymin=-20,
ymax=20,
ytick={-18,-12,...,18},
ylabel style={font=\color{white!15!black},font=\Large},
ylabel={Angle [deg]},
axis background/.style={fill=white}
]
\addplot [forget plot] graphics [xmin=0.5, xmax=11.9, ymin=-20, ymax=20] {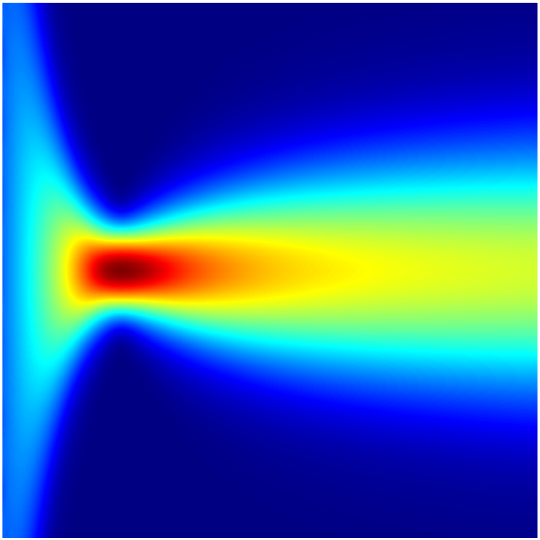};
\end{axis}
\end{tikzpicture}%

\hfill

\begin{tikzpicture}[auto,rotate=0,transform shape]
\begin{axis}[%
width=4.4in,
height=4.4in,
at={(0in,0in)},
scale only axis,
axis on top,
xmin=0.5,
xmax=11.9,
xlabel style={font=\color{white!15!black},font=\Large},
xlabel={Range [m]},
xtick={1,3,...,11},
y dir=reverse,
ymin=-20,
ymax=20,
ytick={-18,-12,...,18},
ylabel style={font=\color{white!15!black},font=\Large},
ylabel={Angle [deg]},
axis background/.style={fill=white}
]
\addplot [forget plot] graphics [xmin=0.5, xmax=11.9, ymin=-20, ymax=20] {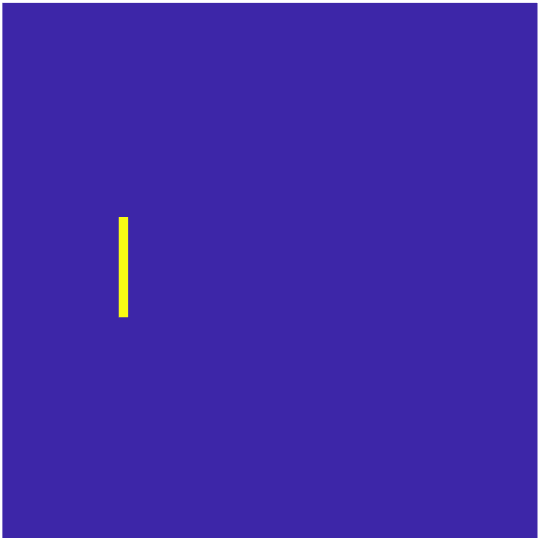};
\end{axis}
\end{tikzpicture}%

%https://texample.net/tikz/examples/swan-wave-model/
}
\caption{\small Examples of beamfocusing codewords and corresponding masks. Note that the masks are defined either to concentrate the radiated energy over extended distances (left) or an extended angular span (right).}
\label{fig:mask_thin}
\end{figure*}

Given the typically large number of columns in $\Am \in \RR^{\Na\times G}$, the program in \eqref{eq:bp_optim_tikhonov2}, which involves matrix inversion in the step obtaining $\Um$ is memory and computationally exhaustive (this is especially true when using CVX solvers). Therefore we solve the beam design problem in \eqref{eq: bp_optim_3} using an iterative method. This method is based on the fact that for fixed $\cv$, $\fv$ can be found as the solution to a linear least-squares problem. Then, for fixed $\fv$ the optimal $\cv$ is equal to a complex number with modulus 1, with its phase equal to  $\angle(\Am^{H} \fv)$. The details of this procedure are provided in Algorithm \ref{alg:beamspot_design}. Two examples of the obtained beamfocusing radiation patterns and the corresponding masks are shown in Fig.~\ref{fig:mask_thin}.

\begin{figure*}[ht]
\centering
\scalebox{0.33}{\includegraphics{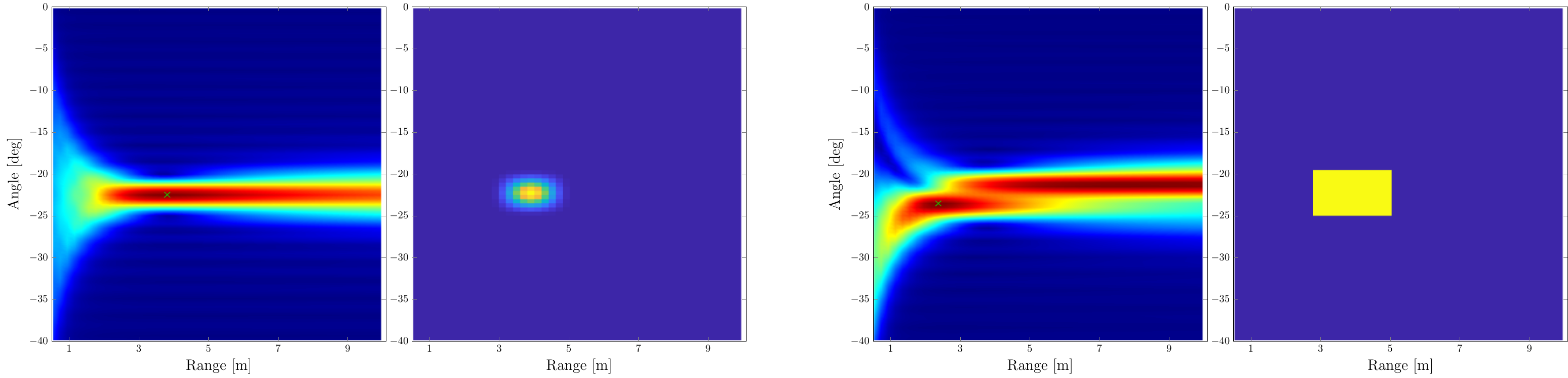}}
\caption{\small Example of beamfocusing codewords derived for (a) Smoothed mask with a Kaiser window with a shape factor of $5.5$, (b) without mask smoothing. Note that in this case, the mask spans a relatively wide angular/distance section as compared to the mask in Fig.~\ref{fig:mask_thin}.}
\label{fig:mask_smoothing}
\end{figure*}

\begin{remark}
    The masks $\bar{\bv} \in \RR^{G}$ as defined above, consist of sharp transitions from the desired spot to the region outside the spot. For a filter design problem, these transitions can lead to instabilities. Moreover, due to the nonlinearity of wave propagation principles, beamfocuing vectors for masks that are wide and long (i.e. large span in angle and range) are difficult to synthesize (compare for example Fig.~\ref{fig:mask_thin} and Fig.~\ref{fig:mask_smoothing}). This is especially true when steering toward the non-boresight directions.  In order to mitigate this issue we propose the use of \textit{windowing} functions on the mask to create a smooth transition and concentrate the energy in a more centralized manner. An example of this can be seen in Fig.~\ref{fig:mask_smoothing} using a 2-D Kaiser window.
\end{remark}

% \begin{figure}[h!]
% \centering
% \includegraphics[scale=0.75, angle=0]{Plots_Figures/beam_spots_scheme.pdf}
% \caption{Schematic representation of a codebook of beamfocusing codewords in the spatial domain. Note that the codewords do not need to uniformly divide the space. As an example, in typical urban deployments, areas with more densely located users can be assigned more refined codewords and vice versa.}
% \label{fig:beam_spots_CB}
% \end{figure}

%%%%%%%%%%%%%%%%%%%%%%%%%%%%%%%%%%%%%%%%%%%%%%%%%%%%%%%%%%%%%%%%%%%%%%%%%%%%%%%%

%%%%%%%%%%%%%%%%%%%%%%%%%%%%%%%%%%%%%%%%%%%%%%%%%%%%%%%%%%%%%%%%%%%%%%%%%%%%%%%%
\section{Simulation Results}\label{simulation_results}
In this section, we present numerical results to evaluate the effectiveness of the proposed schemes. The results in this section are based on the scenario and topology in Fig.\ref{fig:scene_topology}, where we consider numerous trajectories, enumerated I-IV, to simulate the different conditions that may be encountered in a real-world deployment. Unless otherwise stated, at each location along the trajectories, we consider an instance of the extended target generated according to the BND in Section~\ref{sec:target_model} and with the dimensions provided Tab.~\ref{tab:System-Parameters}. Note that the grids defined for parameter estimation do not coincide with the grid used in Section~\ref{sec:target_model} and Tab.~\ref{tab:target-Parameters} to simulate the BND extended target.
%%%%%%%%%%%%%%%%%%%%%%%%%%%%%%%%%%%%%%%%%%%%%%%%%%%%%%%%%%%%%%%%%%%%%%%%%%%%%%
\subsection{Far-field estimation performance (Trajectory I)}
\begin{figure}[h!]
\centering
\includegraphics[scale=0.68, angle=-0]{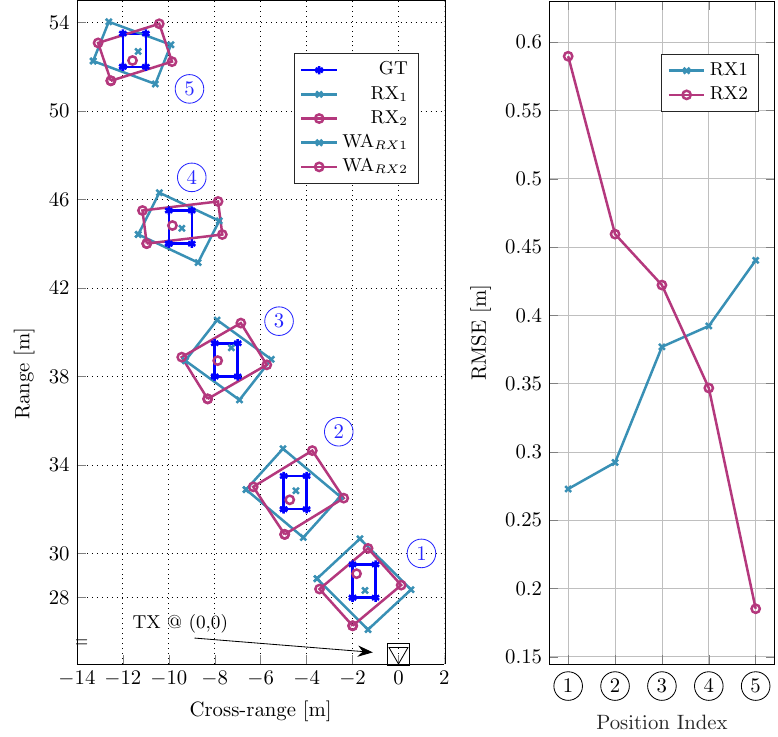}
\caption{Far-field estimates of the extended target along trajectory I in Fig.~\ref{fig:scene_topology}.}
\label{fig:FF_est}
\end{figure}

Referring to Fig.~\ref{fig:scene_topology}, the extended target moving along Trajectory I
is located entirely in the \ac{FF} of the \ac{Tx} and both \ac{Rx} arrays. Figure ~\ref{fig:FF_est} shows the estimated spatial parameters of the extended target. At each step, the extended target is generated 100 times in Monte-carlo (MC) fashion, independently for each of the \ac{Rx} units. The boxes indicate all the point estimates that pass the \ac{OS-CFAR} threshold and have an amplitude within $3$dB of the main peak, local to each \ac{Rx}. These bounding boxes are obtained as the minimum box that fits the convex hull contacting the above-mentioned detection points.  Since the main peak from the \ac{ML} metric in the angle/range domain is wide and leaks into the neighboring bins, the points lying on this main peak also pass the CFAR and lead to enlarged boxes. The system parameters considered in the simulations lead to a bi-static range resolution of approx. $2.34$ [m] and angular resolution of approx. $2.2^\circ$, which are clearly inadequate to resolve the individual scatters in the simulated target box in most cases. Therefore one can calculate a weighted estimate of the detected points after thresholding where the weight of each estimated point corresponds to the value from evaluating \eqref{eq:RxU_ml_estimate_ff}. Then the weighted average estimate of the target is obtained and plotted in Fig.~\ref{fig:FF_est} as the crosses and circles. From this, it can be observed that the estimated target location is very accurate. Note that up to an extent, super-resolution techniques (e.g. sub-space-based methods) can be used alternatively, however, this is out of scope for this paper.  To further demonstrate the accuracy of the \ac{FF} estimation, we calculate the \ac{RMSE} between the \textbf{main} peak at each iteration of the MC process and the center of the box (target). This is plotted in Fig.~\ref{fig:FF_est}(b).      
%%%%%%%%%%%%%%%%%%%%%%%%%%%%%%%%%%%%%%%%%%%%%%%%%%%%%%%%%%%%%%%%%%%%%%%%%%%%%%
\subsection{Multi-static spatial diversity gain (Trajectory IV)}
\begin{figure}[ht]
\centering
\includegraphics[scale=0.67, angle=0]{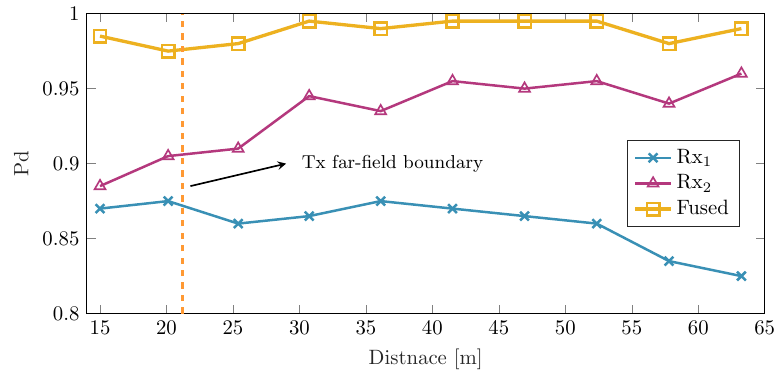}
\caption{Probability of detection for an extended target along trajectory IV in Fig.~\ref{fig:scene_topology}. The x-axis indicates radial distance from the Tx.}
\label{fig:PD_FF_NF}
\end{figure}

In this setup, we aim to demonstrate the effectiveness of multi-static configurations. More specifically, it is well known that under different aspect angles under which targets are observed by the \ac{Rx} units in multi-static configurations, targets can exhibit very different reflection characteristics \cite{chernyak}. To this end, having multiple \acp{Rx} increases robustness in terms of detection. Given that an important distinction of radar systems utilizing multiple radars is to incorporate some level of data fusion between the measurement of individual sensors, we perform the following experiment to demonstrate this. Figure \ref{fig:PD_FF_NF} depicts the average detection probability of the considered extended target at each \ac{Rx}, simulated 200 times for each \ac{Rx} at the given points along Trajectory IV in Fig.~\ref{fig:scene_topology}, where the detection is performed locally at each \ac{Rx} using the \ac{OS-CFAR} thresholding technique. A detection is declared if the main peak passes the threshold and the estimated position resulting from the peak value satisfies $\sqrt{(\hat{x}_p-x_p)^2 + (\hat{y}_p-y_p)^2} < 1$ [m]. Important to note that, each realization is performed according to the BND. This means that in some instances a \ac{Rx} may observe very few or even zero scattering points from the extended target. The \textit{Fused} curve, shows the average detection probability if at each step, either one of the \acp{Rx} has detected the target (i.e. OR operation). This fusion can be performed at a central node.      
%%%%%%%%%%%%%%%%%%%%%%%%%%%%%%%%%%%%%%%%%%%%%%%%%%%%%%%%%%%%%%%%%%%%%%%%%%%%%%
\subsection{Estimation of targets in receiver \ac{NF} (Trajectory III)}
\begin{figure*}[h!]
%\centering
\begin{subfigure}{0.45\textwidth}
\scalebox{0.55}{\includegraphics{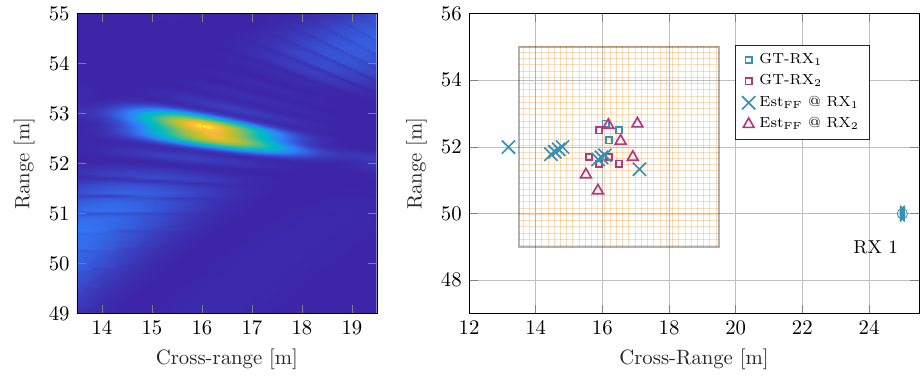}}
\subcaption{Trajectory III-position 2 }
\label{fig:NF_MLE_p1} 
\end{subfigure}
%\hfill
\hspace{0.45cm}
\begin{subfigure}{0.45\textwidth}
\scalebox{0.55}{\includegraphics{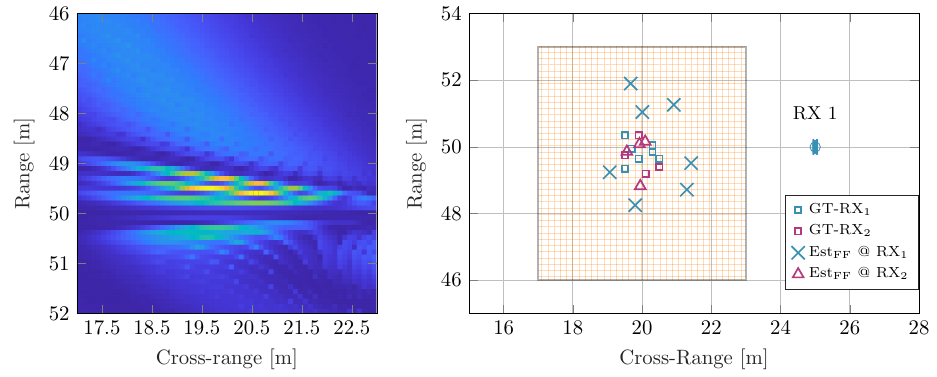}}
\subcaption{Trajectory III-position 1}
\label{fig:NF_MLE_p2} 
\end{subfigure}
\caption{\small The figure shows an extended target located in the \ac{NF} of RX$_1$ (Trajectory III of Fig.~\ref{fig:scene_topology}). After performing an \ac{ML} estimation based on \ac{FF}-bi-static assumptions, the estimated scattering points from RX$_1$ indicate the target to be in the \ac{NF} of this RX unit. Then, a fine-grained search grid is defined in the indicated \ac{RoI}, where the \ac{ML} metric in \eqref{eq:RxU_ml_estimate_opt} is evaluated as shown on the left for each target.}\label{fig:NF_MLE}
\end{figure*}

In Fig.~\ref{fig:scene_topology}, the extended target moving along Trajectory III
is initially estimated to lie in the \ac{NF} of \ac{Rx}$_1$. Figure \ref{fig:NF_MLE} shows the estimates obtained from using the \ac{FF} mismatched model (evaluating \eqref{eq:RxU_ml_estimate_ff}) at the first stage. Then by defining an \ac{RoI} based on these estimates, we evaluate \eqref{eq:RxU_ml_estimate_opt} on a fine-grained grid defined over the \ac{RoI}, comprsing a $6\times6$ [m$^2$] square with $0.05$ [m] pixels. Figure \ref{fig:NF_MLE_p1} corresponds to the closer target location on the trajectory, where it can be observed that it is possible to individually resolve many of the point scatterers with high accuracy (much superior to the bandwidth imposed range resolution).  Figure \ref{fig:NF_MLE_p2} depicts the same procedure, however since the target is now further from Rx$_1$, the individual scattering points can not be resolved. Nonetheless, the estimated location is highly accurate.  Note that the estimates obtained by \ac{Rx}$_2$ for both positions correspond to the \ac{FF} model and since the target lies in its \ac{FF}, the estimates are accurate within the system limits (i.e. range and angular resolution).  

%%%%%%%%%%%%%%%%%%%%%%%%%%%%%%%%%%%%%%%%%%%%%%%%%%%%%%%%%%%%%%%%%%%%%%%%%%%%%%
\subsection{Spectral efficiency enhancement with beamfocusing (Trajectory II)}\label{beamfocusing_results}

By considering a \ac{LoS} channel with free-space path loss ${\rm PL} = (4\pi r/\lambda)^2$, the matching gain $G_{\rm M}$ and achievable spectral efficiency are respectively given by
\begin{subequations}
\begin{align}
 G_{\rm M} =& ~|\av^\herm(r_0,\phi_0)\fv_{\rm X}(\hat{r},\hat{\phi})|^{}\label{eq:match_gain}\\
     {\rm SE} =&  \log_{2}  \left( 1 + \frac{P_{\rm tx}|G_{\rm M}|^{2}}{N_{\rm 0}W}\left(\frac{\lambda}{4\pi r_0}\right)^2 \right), \label{eq:beta_nu}
\end{align}
\end{subequations}
% \begin{align}
%     {\rm SE} = \log_{2}\left(1 + \frac{P_{\rm tx}|\av^\herm(r_0,\phi_0)\fv_{\rm X}(\hat{r},\hat{\phi})|^{2}}{N_{\rm 0}W}\left(\frac{\lambda}{4\pi r_0}\right)^2\right),
% \end{align}
with $(r_0,\phi_0)$ being the true location of the target, and the parameter values specified in Tab.~\ref{tab:System-Parameters}. $\fv_{\rm X} \in \{\fv_{\rm N}, \fv_{\rm F}\}$ is the \ac{Tx} beamforming (/beamfocusing) vector where $\fv_{\rm F}(\hat{\phi})$ is a codeword chosen from a discrete \textit{Fourier} codebook $\in \CC^{\Na\times\Na}$, as the codeword with the closest angular distance of the mainlobe peak with respect to the estimated \ac{AoD} $\hat{\phi}$.

\begin{figure}[ht]
\centering
\includegraphics[scale=0.67, angle=0]{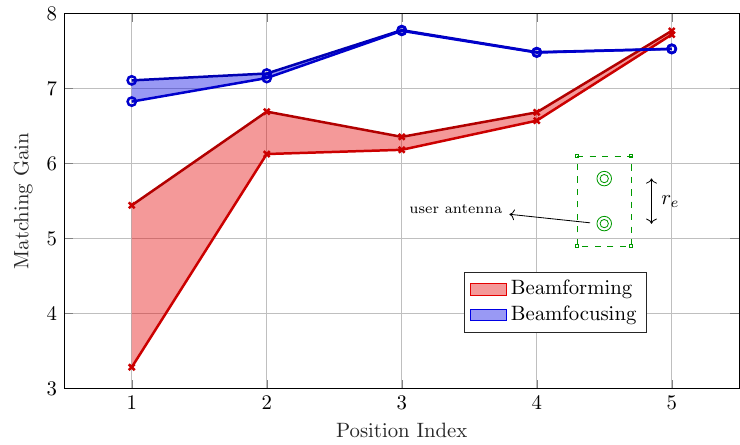}
\caption{Matching gain from beamfocusing and beamforming an extended target along trajectory II in Fig.~\ref{fig:scene_topology}, where the target locations are indexed $1-5$. The filled area shows the gain within a mismatched antenna distance $r_e = 1$ [m] on the extended object.}
\label{fig:SE_NF_e}
\end{figure}

\begin{figure}[ht]
\centering
\includegraphics[scale=0.67, angle=0]{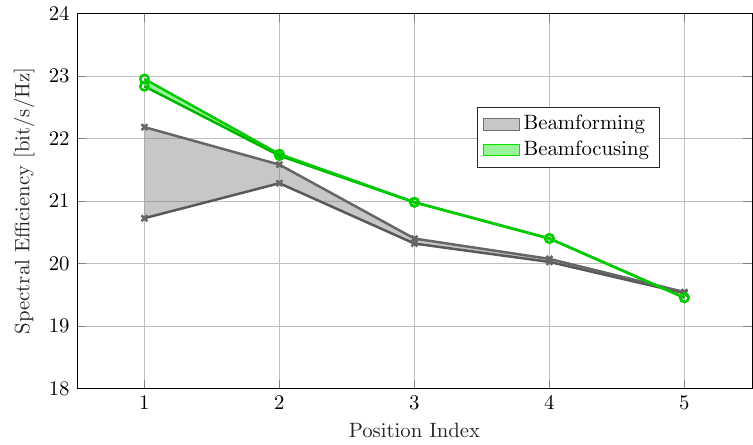}
\caption{Spectral efficiency from beamfocusing and beamforming an extended target along trajectory II in Fig.~\ref{fig:scene_topology}, where the target locations are indexed $1-5$. The filled area shows the gain within a mismatched antenna distance $r_e = 1$ [m] on the extended object.}
\label{fig:SE_NF_e2}
\end{figure}

\begin{figure}[ht]
\centering
\includegraphics[scale=0.67, angle=0]{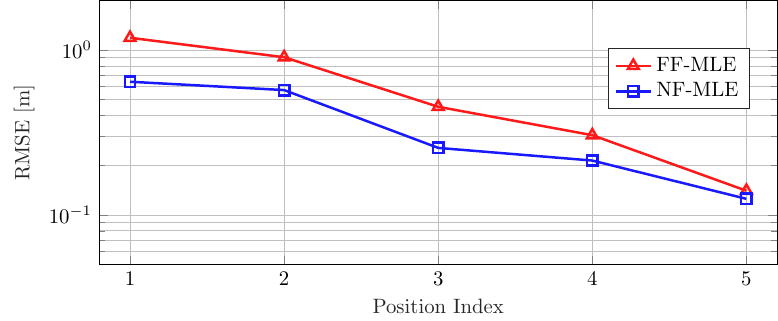}
\caption{Comparison of the estimation error of the target position using a \ac{NF} codeword and \ac{FF} codeword.}
\label{fig:RMSE_NF}
\end{figure}

$\fv_{\rm N}(\hat{r},\hat{\phi})$ is a codeword from the \textit{custom-designed} codebook of beamfocusing vectors, as explained in Section~\ref{BeamFocusing}. Figures \ref{fig:SE_NF_e}, \ref{fig:SE_NF_e2}, show the matching gain and achievable \ac{SE} for the extended target along Trajectory II, respectively. The plots are obtained by calculating the matching gain and \ac{SE} at a hypothetical user antenna that is mounted on the extended target. Since the custom beamfocusing vectors are designed to cover an extended area with a constant gain as in Fig.~\ref{fig:beam_spots_CB}, if the antenna position deviates from the expected location, the Fourier codeword performs significantly worse than beamfocusing words.  To demonstrate the parameter estimation performance with beamfocusing and beamforming more explicitly, we consider a single-point scatterer at each of the locations along Trajectory II and perform parameter estimation (at \ac{Rx}) using beamformed and beam-focused transmission. The results in Fig.~\ref{fig:RMSE_NF} indicate an improved parameter estimation which can be attributed to the increase in \ac{SNR} at the target location.

% \begin{figure}[ht]
% \centering
% \includegraphics[scale=0.67, angle=0]{Plots_Figures/NF_SE_Comp.pdf}
% \caption{Matching gain from beamfocusing and beamforming an extended target along trajectory II in Fig.~\ref{fig:scene_topology}, where the target locations are indexed $1-5$.}
% \label{fig:SE_NF}
% \end{figure}

%%%%%%%%%%%%%%%%%%%%%%%%%%%%%%%%%%%%%%%%%%%%%%%%%%%%%%%%%%%%%%%%%%%%%%%%%%%%%%%%
\begin{table}
	\caption{System parameters}
	\centering
	\scalebox{0.99}{
	\begin{tabular}{|c|c|}
		\hline
		$N=100$ & $M=6$ \\ \hline
		$f_\mathrm{c}=28.0$ [GHz] & $W=128$ [MHz] \\ \hline
% 			$\nu_{\mathrm{grid}}\simeq 24.4$ [KHz] $\equiv$ 40.6 [m/s] & $\tau_{\mathrm{grid}}= 1e-8$ [s] $\equiv$ 6 [m] \\ \hline
% 		$v_{\mathrm{res}}\simeq440$ [km/h] & $r_{\mathrm{res}}\simeq1$ [m]\\ \hline
% 		$v_{\mathrm{max}}=N\cdot v_{\mathrm{res}}$ & $r_{\mathrm{max}}=M\cdot r_{\mathrm{res}}$ \\ \hline
	 $\Pav=26$ [dBm] & $\sigma_{\mathrm{rcs}}=1$ [m$^2$], ~$P_{\rm fa}=10^{-3}$ \\ \hline
	Noise Figure (NF) $=3$ [dB] & Noise PSD  = $2\cdot10^{-21}$ [W/Hz] \\ \hline
	$\Na=64$ & $\Nrf=8$, $B=1$ \\ \hline
	\end{tabular}
	}
	\label{tab:System-Parameters}
\end{table}

\begin{table}
	\caption{Target parameters (see Fig.\ref{fig:PPP_target_model})}
	\centering
	\scalebox{0.99}{
	\begin{tabular}{|c|c|}
		\hline
		Max. length T$_l = 1.5$ [m] & Max. width T$_w = 1$ [m] \\ \hline
		BNM parameter $q \approx 0.01$  & Grid size T$_g = 0.1$ [m] \\ \hline
	\end{tabular}
	}
	\label{tab:target-Parameters}
\end{table}
%%%%%%%%%%%%%%%%%%%%%%%%%%%%%%%%%%%%%%%%%%%%%%%%%%%%%%%%%%%%%%%%%%%%%%%%%%%%%%%%
% \section{ToDo's}
% \begin{itemize}
%     \item Define how to get orthogonality between Txs $\checkmark$
%     \item recheck the geometrical relations $\checkmark$
%     %\item Include Spatio-Temporal model $\checkmark$
%     \item include the generalized near/far field model $\checkmark$
%     \item agree on a parameter estimation method (i.e ML, Sub-space based, etc)
%     \item define a goodness of estimation metric 
%     \item possibly some communication performance metric, especially in the NF case $\checkmark$
%     \item define some simulation scenarios $\checkmark$
% \end{itemize} 

%%%%%%%%%%%%%%%%%%%%%%%%%%%%%%%%%%%%%%%%%%%%%%%%%%%%%%%%%%%%%%%%%%%%%%%%%%%%%%%%
\section{Conclusion}\label{sec:conclusion}
In this work, we propose a two-stage parameter estimation framework for \ac{ISAC} in a multi-static configuration. The framework performs beamforming and  \ac{ML} parameter estimation based on the \ac{FF} assumption in the first stage. The second stage of the scheme is deployed when the estimation results of the first stage indicate the sensing target to be located in the \ac{NF} of the arrays. In particular, if the target is located in the \ac{NF} of the Rx arrays, a high-dimensional \ac{ML} parameter estimation based on the exact signal model (i.e. exact array response model) is carried out in a defined region of interest. If the target is determined to be in the \ac{NF} of the Tx array, the scheme selects beamfocusing codewords, which are designed via solving a magnitude least squares problem, to illuminate the target and the Rxs re-estimate the target parameters.    

%This work proposes a \ac{ML}-based parameter estimation framework for a \ac{mmWave} \ac{ISAC} system in a multi-static configuration using energy-efficient hybrid digital-analog arrays. Due to the typically large arrays deployed in the higher frequency bands to mitigate isotropic path loss, such arrays may operate in the near-field regime. The proposed parameter estimation in this work consists of a two-stage estimation process, where the first stage is based on far-field assumptions, and is used to obtain a first estimate of the target parameters. In cases where the target is determined to be in the near-field on the arrays, a second estimation based on near-field assumptions is carried out to obtain more accurate estimates. In particular, we select \textit{beamfocusing} weights for this array that are designed to have a constant gain over an extended spatial region and re-estimate the target parameters at the receivers. We evaluate the effectiveness of the proposed framework in numerous scenarios through numerical simulations and demonstrate the impact of the custom-designed flat-gain beamfocusing codewords in increasing the communication performance of the system.  

% \begin{figure*}[ht]
% \centering
% \includegraphics[scale=0.75, angle=0]{Plots_Figures/scene_with_NF_TX.pdf}
% \caption{Scenario and target estimates.}
% \label{fig:box_plot}
% \end{figure*}

\section*{Acknowledgment}
The authors would like to acknowledge the financial support by the Federal Ministry of Education and Research of Germany in the program of “Souverän. Digital. Vernetzt.” Joint project 6G-RIC, project identification number: 16KISK030, and by the European Union under the Italian National Recovery and Resilience Plan (NRRP) of NextGenerationEU, partnership on “Telecommunications of the Future” (PE00000001 - program “RESTART”). 
\appendices
\section{Derivation of ML estimate}\label{chap:app_mle}
Here we derive the \ac{ML} estimate for a single scattering point $p$. For ease of notation, we drop some of the arguments. The likelihood-function of $\underline{\yv}_{b}$ in \eqref{eq:rx_blocked} is :  
\begin{align} \label{eq:RxU_likelihood}
    L(\underline{\yv}_{b}; (\varepsilon_p,\nu_p,r_p,\gamma_p,\theta_p, \phi_p)) = \frac{1}{\det(2\pi\sigma^2\Id_{NM\Nrf})^{1/2}} \cdot \nonumber \\
	\exp\left(-\frac{1}{2\sigma^2}\left( (\underline{\yv}_{b} - \varepsilon \Gm_b \underline{\xv}_{b})^\herm (\underline{\yv}_{b}- \varepsilon\Gm_b \underline{\xv}_{b})  \right) \right).
\end{align}
Accumulating all observations up to the $B$-th slot, $\underline{\yv}^{(B)} =[\underline{\yv}_{1}, \dots, \underline{\yv}_{B}]$, the log-likelihood function is given by:
\begin{align} \label{eq:RxU_log_likelihood}
    \ell ( \underline{\yv}^{(B)}; & (\varepsilon_p,\nu_p,r_p,\gamma_p,\theta_p, \phi_p) ) \nonumber \\ & = \log \left( \underline{\yv}^{(B)}; (\varepsilon_p,\nu_p,r_p,\gamma_p,\theta_p, \phi_p)\right) \nonumber \\
    &= \sum_{b=1}^{B} \log \left( L(\underline{\yv}_{b}; (\varepsilon_p,\nu_p,r_p,\gamma_p,\theta_p, \phi_p)) \right).
\end{align}
Using the \ac{ML} estimation procedure for unknown parameters described in \cite{scharf1991statistical} 
and using \eqref{eq:RxU_log_likelihood}, the \ac{ML} estimate in the $b$-th block is given
\begin{align} 
(\hat{\varepsilon}_p,&\hat{\nu}_p, \hat{r}_p, \hat{\gamma}_p,\hat{\theta}_p, \hat{\phi}_p)=  \underset{\varepsilon_p,\nu, r,\gamma,\theta, \phi}{\rm argmax} ~\ell(\underline{\yv}^{(B)}; (\varepsilon_p,\nu_p,r_p,\gamma_p,\theta_p, \phi_p)) \nonumber \\
	&= \underset{\varepsilon_p,\nu, r,\gamma,\theta, \phi}{\rm argmin} \sum_{b=1}^{B} \left\lVert \underline{\yv}_{b} - \varepsilon_{p} \Gm_b(\nu_{p}, r_p,\gamma_{p}, \theta_p,\phi_{p}) \underline{\xv}_b \right\rVert_2^2 \nonumber \\
 	&=  \underset{\varepsilon_p,\nu, r,\gamma,\theta, \phi}{\rm argmin} \sum_{b=1}^{B} \left\lvert \varepsilon_p \right\rvert^2 \underline{\xv}_b^\herm\Gm_b^\herm\Gm_b \underline{\xv}_b - 2\Re\left\{(\varepsilon_p)^* \underline{\xv}_b^\herm\Gm_b^\herm\underline{\yv}_{b} \right\}\;.
	\label{eq:ml_estimate_ue}
\end{align}
defining:  
\begin{align}\nonumber
    \bar{\Um}_{(B)} &= \sum_{b=1}^{B} \lVert \underline{\xv}_{b}\rVert_2^2 \Um_b \Um_b^\herm\\
	\xiv_{(B)}(r,\gamma,\nu) &= \left[ \sum_{b=1}^{B} \underline{\xv}_{b}^\transp \Tm(r, \gamma, \nu) \Ym_b^\herm \Um_b^\herm\right] \nonumber,
\end{align}
Denoting $\Ym_b = [\yv_b[0,0], \dots, \yv_b[N-1,M-1]] \in \CC^{\Na \times NM}$,  and using 
\begin{align}
	\Gm_b^\herm \Gm_b &= 
	\left(\Tm \otimes \Um^\herm_b \bv \av^\herm\fv\right)^\herm \left(\Tm \otimes \Um^\herm_b \bv \av^\herm\fv \right) \nonumber \\
	% &= \left(\Tm^\herm(\tau_0,\nu_0) \otimes \bv^\herm(\phi) \Vm_b \right) \left(\Tm(\tau_0,\nu_0) \otimes \Vm^\herm_b \bv(\phi) \right) \nonumber \\
	&\stackrel{(a)}= \left(\Tm^\herm \Tm \otimes \fv^\herm \av \bv^\herm\Um_{b}\Um_{b}^\herm  \bv\av^\herm \fv \right) \nonumber \\
	% &= \left(\Id_{NM} \otimes \bv^\herm(\phi) \Vm_b \Vm^\herm_b  \bv(\phi) \right) \nonumber \\
	&= \fv^\herm \av \bv^\herm\Um_{b}\Um_{b}^\herm  \bv\av^\herm \fv ~\Id_{NM} \label{eq:app_mle_1} \\
	\intertext{and} 
	\Gm_b^\herm \yv_b &= \left(\Tm \otimes \Um^\herm_b \bv \av^\herm\fv\right)^\herm \yv_b \nonumber\nonumber \\
	&= \left(\Tm^\herm \otimes \fv^\herm \av \bv^\herm\Um_{b}\right) \yv_b \nonumber\\
	&\stackrel{(b)}= \text{vec} \left( \fv^\herm \av \bv^\herm\Um_{b} \text{vec}^{-1}(\yv_b) \Tm^\herm\right) \nonumber \\
	&\stackrel{(c)}= \left( \fv^\herm \av \bv^\herm\Um_{b} \Ym_b \Tm^\herm \right)^\T \nonumber \\
	&= \Tm^\herm\Ym_b^\T \Um^\T_b \bv^*\av^\T\fv^*, \label{eq:app_mle_2}
\end{align}
where $(a)$ stems from mixed property of the Kronecker product: $(\Am \otimes \Bm)(\Cm \otimes \Dm) = (\Am\Cm \otimes \Bm\Dm)$, $(b)$ from the mixed Kronecker matrix-vector product: $(\Am \otimes \Bm)\vv = \vec(\Bm \Vm \Am^\T)$, where $\Vm = \text{vec}^{-1}(\vv)$, and $(c)$ $\text{vec}(\cdot)$-operator applied to row-vector is equivalent to transposing.

For a given $(\nu_p,r_p,\gamma_p,\theta_p, \phi_p)$ tuple, the value $\varepsilon'$ which maximizes \eqref{eq:ml_estimate_ue} is the maximum likelihood estimate of the channel coefficient given that a scattering point is present at $(\nu_p,r_p,\gamma_p,\theta_p, \phi_p)$, which can be evaluated in close form and is given by
\begin{align} \label{eq:mle_ue_g_opt}
    \varepsilon_p' = \frac{ \fv^\herm \av \bv^\herm \xiv_{(B)}(r, \gamma,\nu)}{\fv^\herm \av \bv^\herm \bar{\Um}_{(B)} \bv\av^\herm \fv},
\end{align}
Substituting expression \eqref{eq:mle_ue_g_opt} into Eq. \eqref{eq:ml_estimate_ue}, the \ac{ML} estimate can then be derived  as:
\begin{align} \label{eq:RxU_ml_estimate_opt}
    (\hat{\nu}_p, \hat{r}_p, \hat{\gamma}_p,\hat{\theta}_p, \hat{\phi}_p) = \underset{\nu, r,\gamma,\theta, \phi}{\rm argmax} \frac{\left\lvert \fv^\herm \av \bv^\herm \xiv_{(B)}(r, \gamma,\nu) \right\rvert^2}{\fv^\herm \av \bv^\herm \bar{\Um}_{(B)} \bv\av^\herm \fv},
\end{align}
where the space is $\Gamma_{\rm ML}\eqdef \RR^{5P}$.

\FloatBarrier
\bibliography{IEEEabrv,book}

\end{document}